\documentstyle[12pt,aaspp4,flushrt]{article}
\newcommand{\etal}{\mbox{ et~al.}}
\newcommand{\eg}{\mbox{e.g.}}
\newcommand{\kms}{\mbox{ km~s$^{-1}$}}

\newcommand{\gapp}{\stackrel{\scriptstyle >}{\scriptstyle \sim}}

\newcommand{\ciii}{CIII$\left.\right]$}

\newcommand{\dm}{\langle \Delta m \rangle}

\begin{document}

\title{Limits on Cosmological Models From \\
     Radio-Selected Gravitational Lenses 
\footnote{Observations reported here were made with the Multiple Mirror 
Telescope Observatory, which is operated jointly by the University of Arizona
and the Smithsonian Institution}
\footnote{Observations reported here were obtained, in part, at 
MDM Observatory,
a consortium of the University of Michigan, Dartmouth College
and the Massachusetts Institute of Technology.}
\footnote{This research made use of the NASA/IPAC Extragalactic Database (NED)
  which is operated by the Jet Propulsion Laboratory, Caltech, under contract
  with the National Aeronautics and Space Administration}
\footnote{We have made use in part of finder chart(s)
         obtained using the Guide Stars Selection System Astrometric Support
         Program developed at the Space Telescope Science Institute (STScI is
         operated by the Association of Universities for Research in Astronomy,
         Inc., for NASA) }
}

\author{ E. E. Falco, C. S. Kochanek \& J. A. Mu\~noz }
\affil{Harvard-Smithsonian Center for Astrophysics \\
 60 Garden Street \\ Cambridge MA 02138}

\begin{abstract}
We are conducting a redshift survey of 177 flat-spectrum radio sources in
3 samples covering the 5~GHz flux ranges 50--100, 100--200 and
200--250 mJy. So far, we have 
measured 124 redshifts with completenesses of 80\%, 68\% and 58\%
for the bright, intermediate, and faint flux ranges. Using the newly
determined redshift distribution we can derive cosmological limits
from the statistics of the 6 gravitational lenses in the JVAS sample of
2500 flat-spectrum radio sources brighter than 200 mJy at 5~GHz.  For
flat cosmological models with a cosmological constant, the limit
using only radio data is $\Omega_0 > 0.27$ at 2$-\sigma$ 
($0.47 < \Omega_0 < 1.38$ at 1$-\sigma$).   The limits are statistically
consistent with those for lensed quasars, and the combined radio + optical
sample requires $\Omega_0 > 0.38$ at 2$-\sigma$ ($0.64  < \Omega_0 < 1.66$ 
at 1$-\sigma$) for our most conservative redshift completeness model 
and assuming that there are no quasar lenses produced by spiral galaxies.
Our best fit model improves by approximately 1$-\sigma$ if extinction
in the early-type galaxies makes the lensed quasars fainter by
$\Delta m = 0.58 \pm 0.45$ mag, but we still find a limit of
$\Omega_0 > 0.26$ at 2$-\sigma$ in flat cosmologies.
The increasing fraction of radio galaxies as compared to quasars at fainter
radio fluxes (rising from $\sim$10\% at 1 Jy to $\sim$50\% at 0.1 Jy) explains
why lensed optical emission is common for radio lenses and partly explains
the red color of radio-selected lenses. 
\end{abstract}

\keywords{cosmology: observations ---  galaxies: distances and redshifts ---
          quasars --- radio galaxies --- gravitational lensing }

\section{Introduction}

The global geometry of the universe, usually specified by its matter 
density $\Omega_0$ and a cosmological constant $\lambda_0$, remains a 
significant source of uncertainty in cosmology.  Current summaries of 
the constraints (\eg, Ostriker \& Steinhardt 1995; Krauss \& Turner 1995) favor
a low matter density ($\Omega_0 \sim 0.3$) either with or without
a cosmological constant.  The expectations of a 
low matter density are driven by observations
of large-scale structure, the cluster baryon fraction, and nucleosynthesis
(\eg, Peacock \& Dodds 1994; White \etal\ 1993; Copi, Schramm \& Turner 1995).
Globular cluster ages, however, no longer require a low $\Omega_0$ due
to the Hipparcos revisions of the distance scale (e.g. Chaboyer \etal\ 1997). 
A flat (inflationary) model would then require a cosmological
constant $\lambda_0 \sim 0.7$.

The number of gravitational lenses found in systematic surveys for
lenses is a strong constraint on cosmological models, particularly
models with a large cosmological constant (Turner 1990; Fukugita,
Futamase \& Kasai 1990).  Quantitative analyses of surveys for
multiply imaged quasars (Kochanek 1993, 1996a; Maoz \& Rix 1993)
currently give a formal two-standard deviation (2$-\sigma$) upper
limit on the cosmological constant in flat models ($\Omega_0+\lambda_0=1$) 
of $\lambda_0 < 0.66$, and the lensing constraints are almost identical
to the very preliminary results using Type Ia supernovae by  
Perlmutter \etal\ (1997).  The statistical
uncertainties are dominated by the Poisson errors from the small
number of lensed quasars and the uncertainties in the local number
counts of galaxies by type.  The limits are also subject to several
systematic errors; the principal ones are extinction 
(e.g. Kochanek 1991, 1996a; Tomita 1995; Malhotra, Rhoads \& Turner 1996;
Perna, Bartelmann \& Loeb 1997), 
galaxy evolution (e.g. Mao 1991; Mao \& Kochanek 
1994; Rix \etal\ 1994), the quasar discovery process (Kochanek 1991),
and the model for the lens galaxies (\eg, Maoz \& Rix 1993; Kochanek 
1993, 1994, 1996a).  

We can eliminate two of these systematic errors, extinction and the
quasar discovery process, by using the statistics of radio-selected
lenses to constrain the cosmological model.  Radio-selected lenses are
immune to extinction in the lens galaxy, and radio lens searches work
from flux-limited surveys that avoid the complicated systematic and
completeness issues of quasar catalogs.  Agreement between the optical
and radio samples is a powerful check on some aspects of the lens
galaxy models and for unanticipated systematic errors due to the large
differences of the two samples in their redshift distributions,
luminosity functions, and fractions of lensed objects.  Moreover, 
we can reduce the Poisson uncertainties by performing a joint analysis
if the samples are statistically consistent. 

Unfortunately, the radio lens surveys use flux limits where there is
little direct information on the source redshift distribution.
Complete redshift surveys exist only for sources brighter than $> 300$
mJy (e.g. the CJI/CJII samples, Henstock \etal\ 1995; the Parkes
Half-Jansky Sample, PHFS, \cite{drink97}; and other Parkes samples,
Peacock \& Wall 1981; Wall \& Peacock 1985; Dunlop et al. 1986, 
1989; Allington-Smith et al. 1991), while the 3 
large radio lens surveys, the MIT-Greenbank Survey (MG, Burke, Leh\'ar \&
Conner 1992), the Jodrell Bank-VLA Astrometric Survey (JVAS, Patnaik
1994; Patnaik \etal\ 1992a; King \etal\ 1996; Browne \etal\ 1997), and the
Cosmic Lens All-Sky Survey (CLASS; Myers \etal\ 1995; 
Browne \etal\ 1997; Jackson \etal\ 1997) 
have flux limits of 50--100 mJy, 200 mJy, and
25--50 mJy respectively.  The typical lens found in a survey is
magnified from still fainter fluxes, typically about 25--50\% of the
survey flux limit.  In Kochanek (1996b) we found that the
uncertainties in the redshift distribution, or equivalently the radio
luminosity function, led to serious systematic uncertainties in the
cosmological limits that could be set using the JVAS survey.  There
was, however, a strong correlation between the mean redshift of the
flat spectrum radio sources with fluxes from 50 to 300 mJy and the
inferred cosmological model (for flat models with a cosmological
constant, the expected mean redshift ranged from $0.4$ for
$\Omega_0=0$, to $1.9$ for $\Omega_0=1$, and to $4.0$ for
$\Omega_0=2$).

The large variation in the average source redshift with cosmological
model means that a modest redshift survey will produce strong
cosmological constraints.  In \S2 we report on the redshift
distribution of three samples of flat-spectrum radio sources in the
flux range from 50 to 250 mJy.  In \S3 we use the new redshift
information to redetermine the limits on cosmological models using
only radio-selected lenses and compare the results to the limits using
lensed quasars and the joint sample.  Finally in \S4 we discuss the
remaining systematic uncertainties and the need for future
observations.

\section{Observations}

The JVAS survey examined 2500 flat-spectrum radio sources with 
($\nu=$ 5~GHz) fluxes brighter than 200 mJy (Patnaik
1994; Patnaik \etal\ 1992a; King \etal\ 1996; Browne \etal\ 1997). 
Because gravitational
lensing magnifies the sources, the typical lensed source in the JVAS
sample originally had a flux between 50 and 200 mJy.  Unfortunately,
the only published redshift survey of flat-spectrum radio sources at
these flux levels contained only 41 sources brighter than 100 mJy with
28 measured redshifts (the Parkes Selected Area Survey, Dunlop 
\etal\ 1989).

To allow us to determine the first limits on the cosmological model
using radio-selected lenses, we first selected three flat-spectrum
samples to cover the flux range of the sources found as lenses in the
JVAS survey (see Tables 1--4).  The first sample of 69 objects was
selected from the faint tail of the JVAS sample to have 5~GHz fluxes
between 200 and 250 mJy.  The second sample of 63 sources was selected
from the MIT-Greenbank (MG) Survey (Burke, Leh\'ar \& Conner 1992)
with fluxes between 100 and 200 mJy.  The third sample of 45 sources
was also selected from the MG Survey with fluxes between 50 and 100
mJy.  Each sample included all sources meeting the flux criterion in a
fixed area of the sky determined by the epoch of the main
spectroscopic observing run.

For each sample we first obtained $I$ band images to obtain an optical
identification and an estimate of the $I-$band flux for each source.
We chose the $I$ band because the faintest radio sources tend to be
red (\eg, Webster \etal\ 1995).  The images were obtained at the Fred
Lawrence Whipple Observatory (FLWO) 48'' telescope and at the MDM
Observatory Hiltner 2.4 m telescope.  At FLWO, the detector was a Loral
2048$^2$ CCD with a Kron-Cousins $I$ filter.  The pixel scale of the
CCD is $0\farcs63$ (binned $2\times2$) the nominal gain is 2.30
electrons/ADU, and the nominal read-out noise is 7.0 electrons per
pixel (unbinned).  At MDM, the detector was a
Tektronix 1024$^2$ CCD, with gain 3.45 electrons/ADU, read-out noise
4.0 electrons per pixel, and pixel scale 0\farcs275.  The exposure
times ranged from 3 to 30 minutes; the identification of each source
was relatively simple, because all the radio sources were selected
from VLA imaging surveys with arcsecond positional accuracy.  The
images were reduced by standard procedures, using the HST Guide Star
Catalog (GSC) to perform the astrometic identifications.  Our
observations were not necessarily obtained under photometric
conditions; therefore, we calibrated the instrumental magnitudes only
approximately, using the magnitudes of GSC stars in our fields, and
assuming a mean $V-I=1.0$ color for these stars.  As a result, our
photometry has significant absolute uncertainties.

\begin{deluxetable}{clccccccc}
\small
\tablewidth{0pt}
\tablecaption{Sample Properties}
\tablehead{ 
\colhead{Sample} &\colhead{Source} &\colhead{Flux} & \colhead{Objects} &
\colhead{Ident.} &\colhead{Det.} & \colhead{Completeness} &
\colhead{$\bar{z}$} &\colhead{$\sigma_z$}\nl
 & &\colhead{(mJy)} & &
 & & \colhead{(\%)} &
 &}
\startdata
 1 &JVAS & 200--250 &  69  &  55  & 12 & 80 &  1.19  &  0.84 \\
 2 &MG   & 100--200 &  63  &  43  & {\phn}6 & 68 &  1.22  &  0.96 \\
 3 &MG   & {\phn}50--100  &  45  &  26  &  {\phn}6 & 58 &  1.28  &  1.08 \\
\hline
\enddata
\label{data}
\end{deluxetable}

We obtained spectra of the objects using the FLWO 60'' Tillinghast
telescope and the FAST spectrograph for the optically brighter
sources, and the MMT and the Blue Channel spectrograph for the fainter
sources.  The useful range of wavelengths is $\sim$ 3200--8600 \AA, with a
resolution of 1.46 (1.96) \AA $\,$ pixel$^{-1}$ for the 60'' (MMT)
spectra.  We used slits of widths 1-2\arcsec, depending on observing
conditions, and a 300-line/mm grating.  The exposure times usually
ranged from 5 to 60 minutes; a small number, the optically faintest
sources, required up to 120 minutes.  We made a single pass through
all the sources with a fixed maximum exposure time, and then used the
remaining time to fill in the redshifts of the fainter sources.
We analyzed 
emission line spectra (mostly quasars, but also a few galaxies) 
with the IRAF task emsao to find their redshifts. We analyzed 
absorption line spectra (early-type galaxies) with the 
IRAF task xcsao and appropriate templates. 

In Tables \ref{s1id}, \ref{s2id} and \ref{s3id} we display the contents  
and our final results for samples 1, 2 and 3 respectively; in 
columns from left to right we list for each object 
its name, right ascension and declination (B1950), $I$ 
magnitude, $I$ magnitude standard error, redshift, redshift standard
error, classification (see below) 
and emission or absorption lines used to classify
each object and compute its redshift. Table \ref{serendip} contains an 
additional 5 optically bright JVAS sources (200--250 mJy) 
for which we obtained redshifts that lay outside
the Sample 1 survey region.

Only a handful of galaxies at $z<0.3$ were clearly distinguishable
from point sources due to the seeing and surface brightness limits in
our photometric observations. Thus, our objects are labeled according
to their spectroscopic classification. We made the following
classifications: E for objects where we detected only absorption lines
usually found in early-type galaxies and L where we also detected
emission lines usually found in late-type galaxies; Q (quasar) for
objects where we detected permitted emission lines with FWHM $\gapp
2000$ \kms\ in their rest frames; and b for BL Lac objects where we
detected only weak absorption lines but no emission lines (the
redshifts are tentative for these objects). We further labeled the
quasars N (for NAL) or B (for BAL) according to the presence of
absorption lines that were significantly narrower or broader,
respectively, than their emission lines (\eg, \cite{an93}).  The
fraction of identified objects depended mainly on the weather
conditions; the lowest completeness was that of Sample 3, where a
third of the run was lost.  Table \ref{data} shows the total number of
objects, the number of measured redshifts, the number of detected
objects (see below), the completeness, the mean redshift $\bar{z}$ and its
standard deviation $\sigma_z$ for each sample. Our samples included a
total of 89 quasars (4 of which were BAL quasars), 33 galaxies and 2
BL Lac objects (see Tables \ref{s1id}, \ref{s2id} and \ref{s3id}).

\begin{figure}
{\epsfxsize=15cm \epsfbox{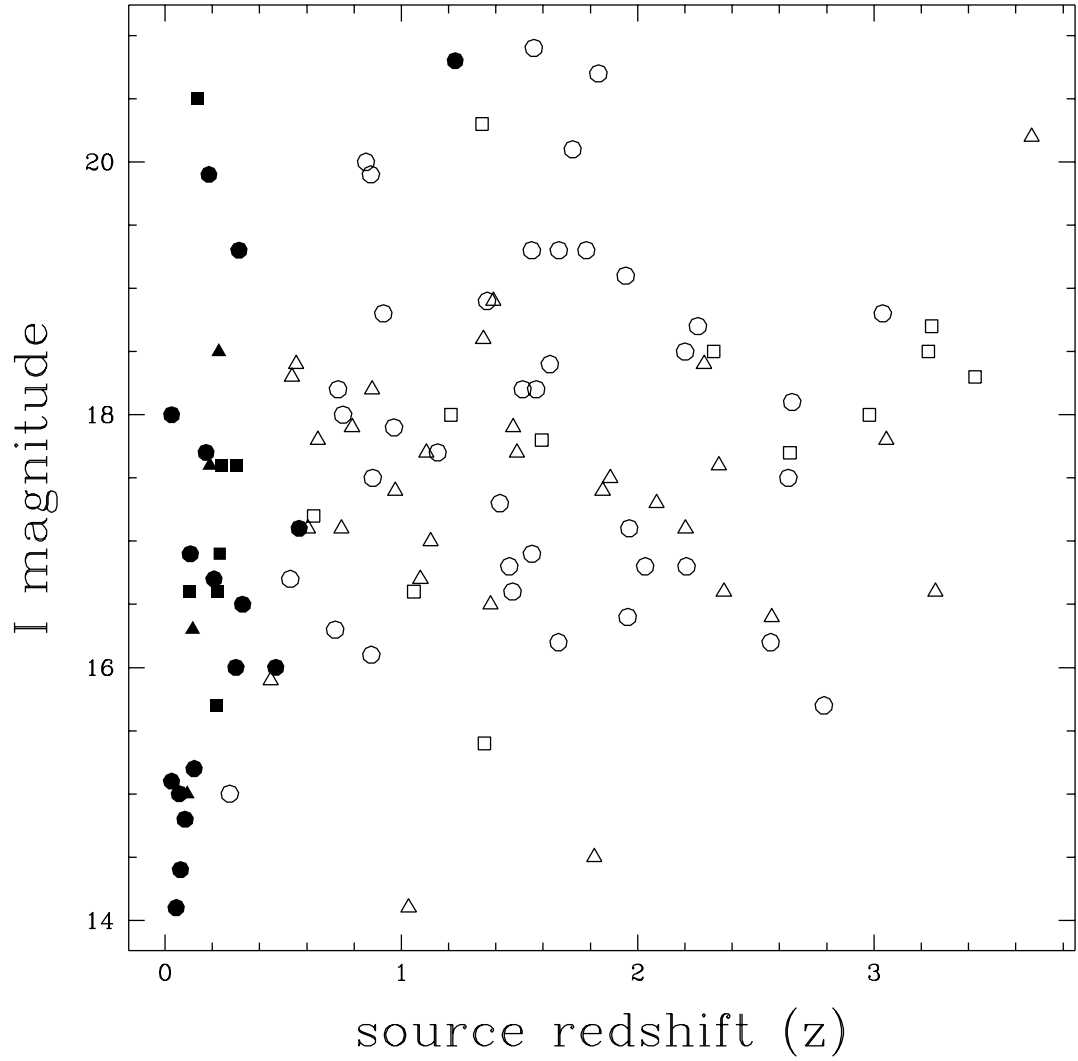}}
\figcaption{Distribution of $I$ magnitudes of detected quasars and galaxies 
as a function of redshift for samples 1, 2 and 3 (circles, triangles and
squares, respectively).  The empty points are quasars and the filled points
are galaxies. }
\label{f1}
\end{figure}

\begin{figure}
{\epsfxsize=15cm \epsfbox{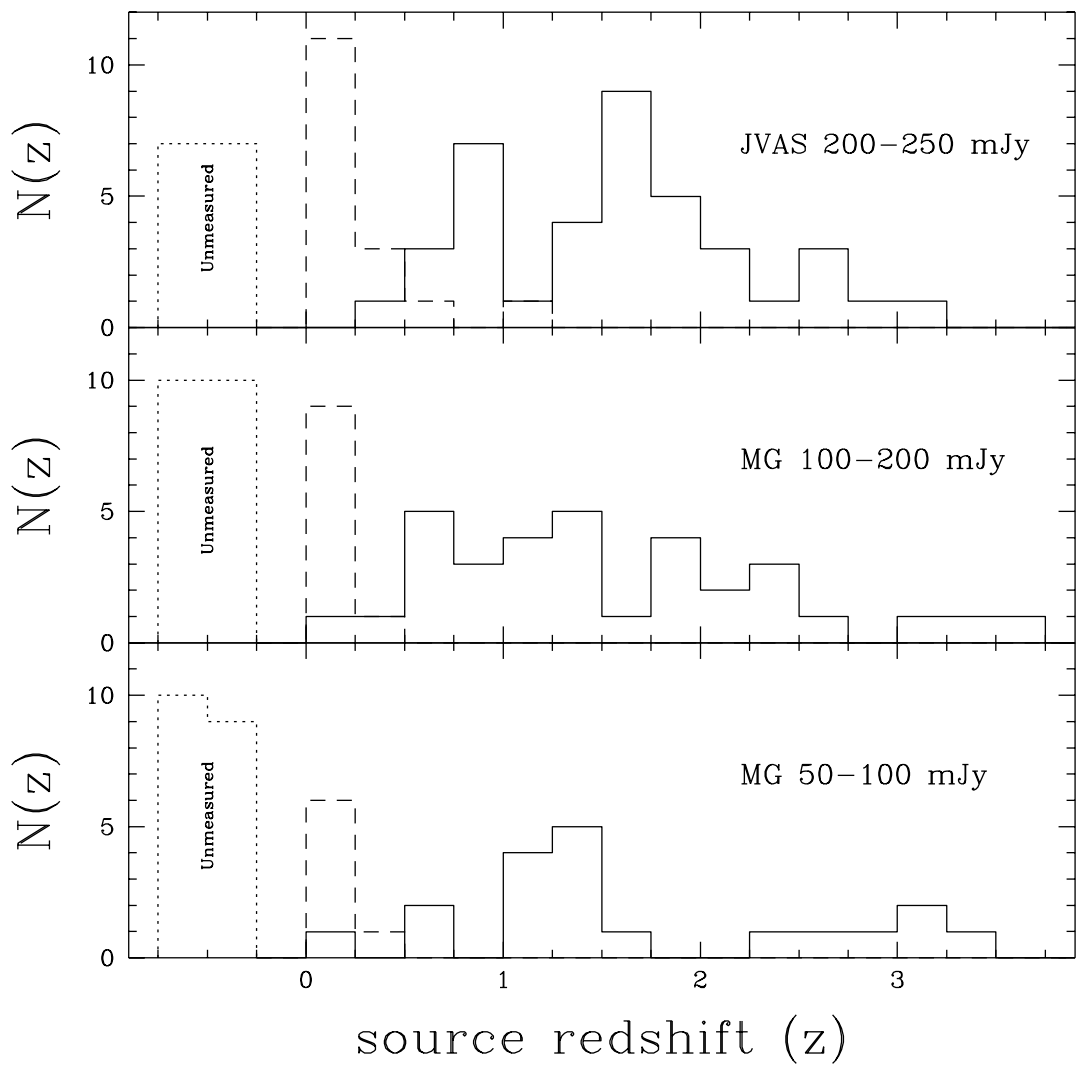}}
\figcaption{Histograms of redshifts for galaxies (dashed) and quasars 
(solid) in  
samples 1--3. The histograms at negative redshifts show the 
numbers of objects with undetermined redshifts. }
\label{f2}
\end{figure}

Figure \ref{f1} shows the $I-$band magnitude distribution as a function of 
redshift for our 3 samples.
Because there is no simple relation between optical magnitude and
redshift that we can use to estimate redshifts, we are forced to use
completeness models in our estimates of the luminosity function.
Figure \ref{f2} shows histograms of the redshifts.
We attempted to acquire 
spectra of almost all the sample objects, because we could easily 
detect emission lines even in the faintest sources.  Thus, we know
that most of the objects lacking redshifts also lack emission lines
and must be early-type galaxies rather than quasars or galaxies
with strong emission lines. We estimate that 
the 24 objects for which we obtained spectra that yielded no redshift 
are early-type galaxies with unknown redshifts. 

\begin{figure}
{\epsfxsize=15cm \epsfbox{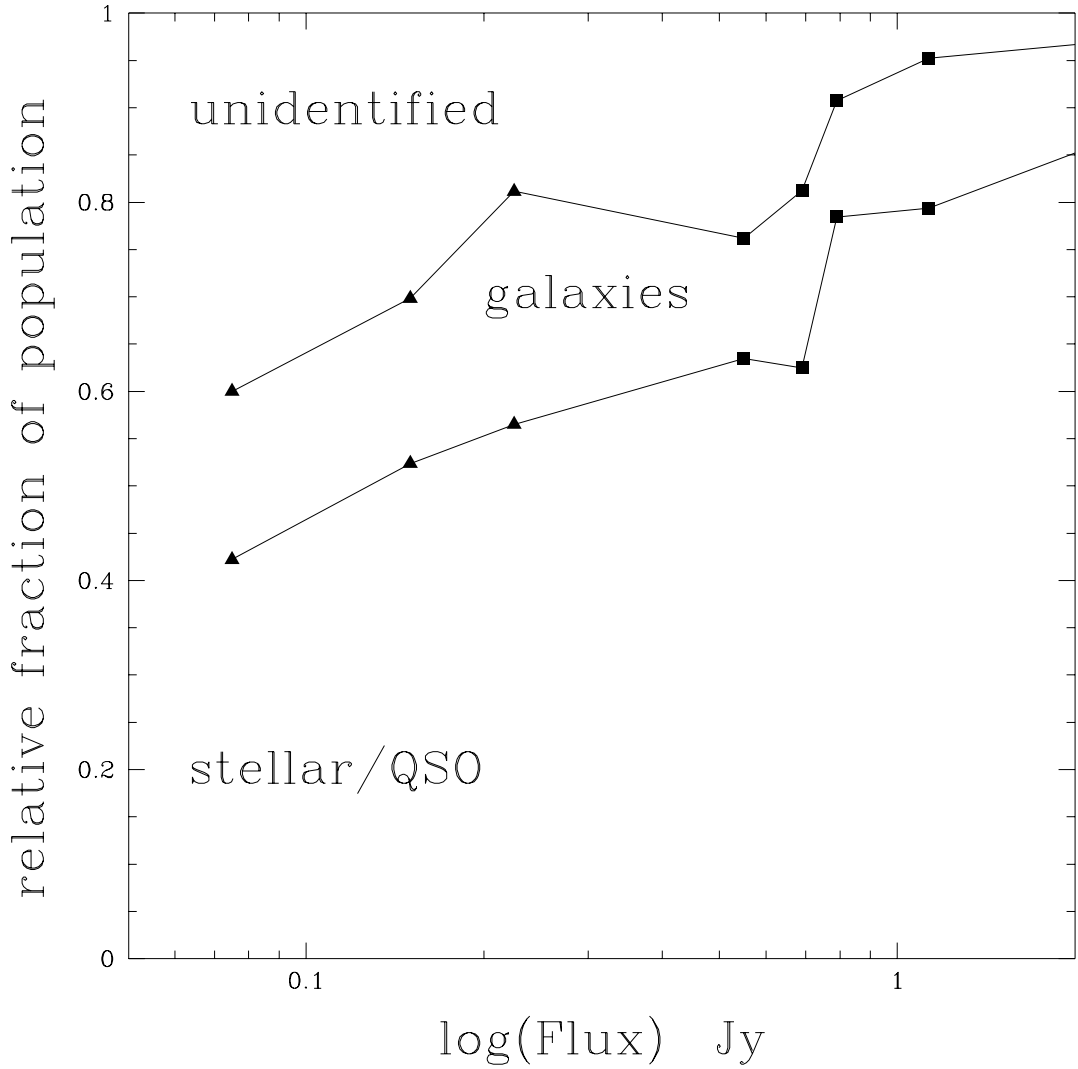}}
\figcaption{Source populations as a function of radio flux for the PHFS 
sample (squares) and our sample (triangles).  
Of 53 unidentified objects, 24 were spectroscopically detected;
thus, we know they lack emission lines. }  
\label{f3}
\end{figure}

One clear trend in the samples as we move to fainter fluxes is the
rapidly increasing proportion of radio galaxies.  In (radio) bright samples
(e.g. Drinkwater \etal\ 1997), the overwhelming majority ($>$90\%) of
the sources are radio quasars, while in our faintest sample we
estimate that $\lesssim$50\% of the sources are quasars.  The trend with
radio flux is illustrated in Figure \ref{f3}.  
The rapid evolution of the population distribution helps to explain
the very different properties of the radio lenses from those of radio sources
at the same observed fluxes (\eg, \cite{mal96}).  
The intrinsically fainter lenses are
likely to be optically extended (as seen in HST observations of
MG~0414+0534 (Falco et al. 1997) and CLASS~1608+656
(Jackson, Nair \& Browne 1997)) and redder than both bright radio
sources and optically-selected quasars.

\begin{figure}
{\epsfxsize=15cm \epsfbox{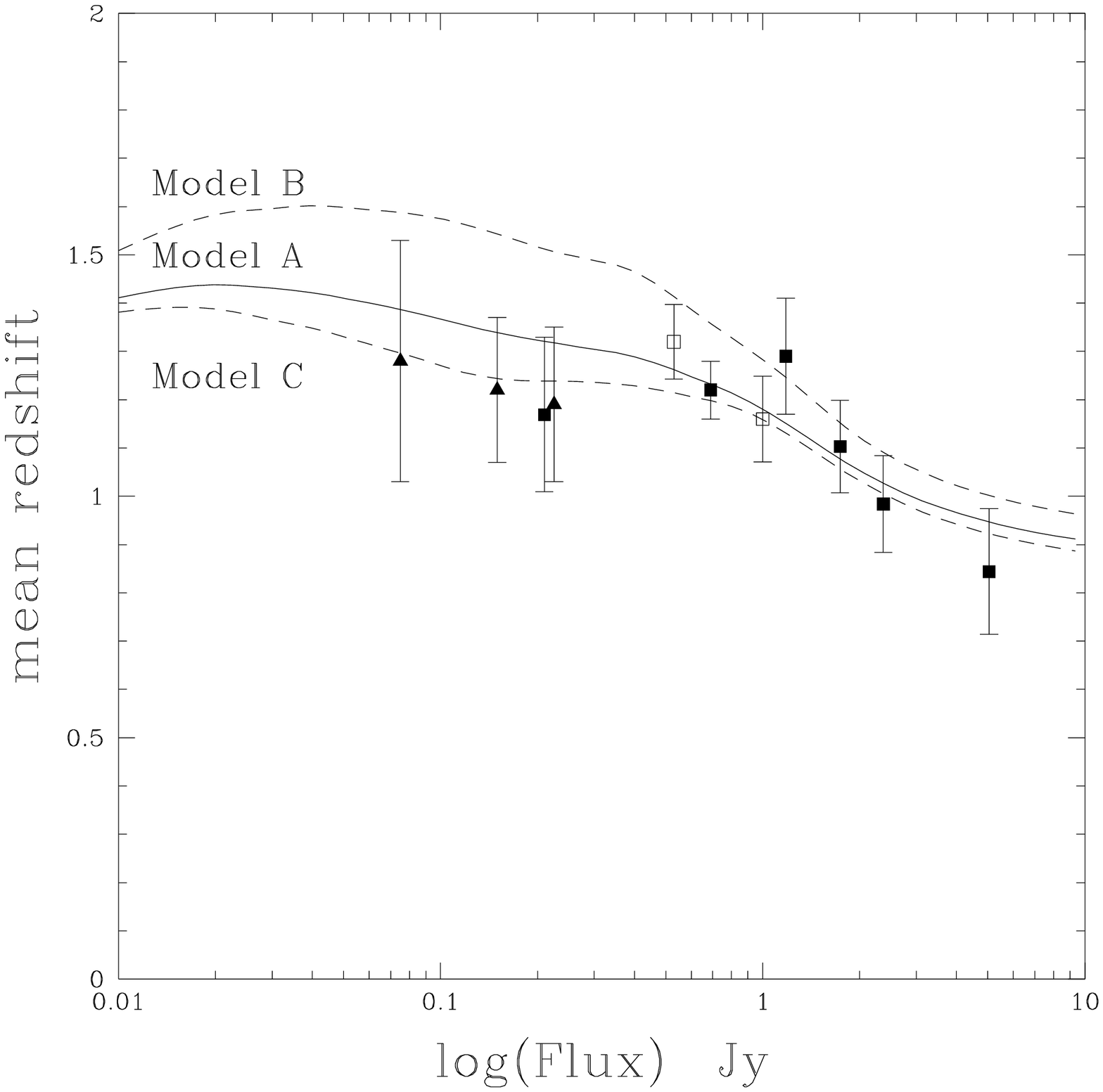}}
\figcaption{Mean source redshift as a function of radio flux.  The points 
  show the observed mean and its standard error for the Parkes samples (solid
  squares), CJI/CJII samples (open squares), and our samples (solid
  triangles).  The lines show the true redshift distributions for Models
  A, B, and C.  In Model A (solid line) the unmeasured redshifts are
  assumed to have the same distribution as the measured redshifts,
  while in Model B the completeness decreases linearly with redshift
  and in Model C it increases linearly with redshift.  Thus, the true
  redshifts are larger than in the observed sample for Model B and
  smaller for Model C. }  
\label{f4}
\end{figure}

\section{Revised Cosmological Limits}

We calculated the expected number of lenses using the techniques for
constructing the radio luminosity function (RLF) outlined in Kochanek
(1996b).  In the analysis we used only the two brighter samples, as
the lower completeness of the third sample would introduce too many
uncertainties.  We used three different completeness models to estimate 
the unmeasured redshifts.  In Model A the unmeasured redshifts have
the same statistical distribution as measured ones, in Model B the
redshift completeness was a linearly declining function of redshift,
and in Model C the redshift completeness was a linearly increasing
function of redshift.  Model B biases the distribution to higher
redshifts, while Model C biases it to lower redshifts.  As illustrated
in Figure \ref{f4}, the effects of the completeness model on the mean
redshifts are modest, particularly when we bias the redshifts
downwards.  To the data used to constrain the RLF in Kochanek (1996b)
we also added the results of the PHFS redshift survey (Drinkwater et al. 1997) 
of 323 flat-spectrum sources brighter than 500 mJy at 2.7~GHz.
Model A has a slightly higher mean redshift than the sample means because 
of the evolution and smoothness constraints required to compensate for
missing data and noise (see Figure \ref{f4}).  The $\chi^2$ of the 
model A--C fits to the binned redshift distributions for samples 1 and 2 are 
statistically acceptable. 

Because current evidence favors the dark matter model of early-type galaxies 
with a singular core (e.g. HST observations of galaxy cores, Byun et al. 1996,
and the previous results of lens statistical studies) we decided to use the
simple singular isothermal sphere (SIS) model for the lens galaxies.
The expected number of lenses changes little if we allow the mass
distributions to
be ellipsoidal rather than spherical or de Vaucouleurs rather than SIS
when the models are normalized to fit the observed distribution of image
separations  (see Kochanek 1996a, 1996b).  
We model the distribution of galaxies 
using Schechter functions for the early-type and the spiral populations,
with constant comoving densities of $n_e = (0.61\pm0.21) h^3 \times
10^{-2} $ Mpc$^{-3}$ and $n_s = (0.79\pm0.26) h^3 \times 10^{-2} $
Mpc$^{-3}$.  The total density of galaxies is more certain than the
division by type, so we restricted $n_e+n_s = (1.40 \pm 0.17) h^3
\times 10^{-2} $ Mpc$^{-3}$.  The overall galaxy density is normalized
as in Loveday \etal\ (1992) while the division by type is taken from
Marzke \etal\ (1994).  The Loveday \etal\ (1992) sample is too deep
for accurate galaxy typing, while the Marzke \etal\ (1994) sample is
too shallow to represent the mean density due to local structures.
Both the spirals and the ellipticals are given the mean  Schechter
function slope of $\alpha_e = \alpha_s = -1.0 \pm 0.1$ of the Marzke et
al. (1994) sample.  Galaxy luminosities are converted to the dark matter
velocity dispersions of the SIS lens model, $\sigma_*$, using
``Faber-Jackson'' relations with $L/L_* =
(\sigma/\sigma_*)^\gamma$.  For the early-type galaxies we adopted
$\gamma_e = 4.0 \pm 0.5$ and $\sigma_{*e} = 225.0\pm22.5$ km s$^{-1}$
based on the models of Kochanek (1994) for the stellar dynamics of
early-type galaxies in singular isothermal halos.  Both parameters are
given uncertainties of approximately twice their formal standard
errors to encompass possible systematic errors.  For the spirals we
adopted the model of Fukugita \& Turner (1990) with $\gamma = 2.6 \pm
0.2$ and $\sigma_{*s} = 145 \pm 10$ km s$^{-1}$. Although the cross sections
of spiral galaxies depend strongly on inclination, their 
inclination-averaged total cross section is still well 
represented by the SIS model (Keeton \& Kochanek 1997).  

Using the methods of Kochanek (1993, 1996ab) we computed the joint
probability of finding the observed number of lenses and fitting their
separations using luminosity functions and models constrained by
Gaussian priors for the measured values (log-normal in the case of the
comoving density).  We computed the likelihoods in the
$\Omega_0$-$\lambda_0$ plane; the increased number of parameters
with the inclusion of the spirals precluded the full Bayesian
calculation used in Kochanek (1996a) because of the need to integrate 
over all the unknown variables.  We instead found the maximum likelihood 
model for each cosmology by optimizing all the other parameters.

We used the same sample of quasar lenses as in Kochanek (1996a), with
862 quasars and 5 lenses (1208+1011, H~1413+117, LBQS~1009--0252,
PG~1115+080, and 0142--100).  For the separation distributions we also
added two additional lensed quasars where we can model the selection
function (BRI~0952--0115 and J03.13), by including the probability
they would have their observed separations given the range of
separations over which they could be detected.  We modeled the JVAS
survey as a sample of 2500 sources with a flux limit of 200 mJy
containing 6 lenses (B~0218+357, MG~0414+0534, B~1030+074, B~1422+231,
B~1938+666 and B~2114+022).  We may be overestimating the lensing rate
by including the 5\% of sources and the lenses (B~1938+666
and possibly B~2114+022) with significant extended radio structure, because 
finite source size or multiple source components 
significantly increase the lensing probability (Kochanek \&
Lawrence 1990).  For the separation distribution we also added the
additional radio lenses where we can model the selection function
(CLASS 0712+472, MG~0751+2716, MG~1131+045, MG~1549+3047, CLASS~1600+434, 
CLASS~1608+656, MG~1654+1346, CLASS~1933+507 and CLASS~2045+265).  
See Keeton \& Kochanek (1996), Browne \etal\ (1997) and Jackson \etal\ (1997)
for a summary of the lenses and their properties.  

\begin{figure}
{\epsfxsize=15cm \epsfbox{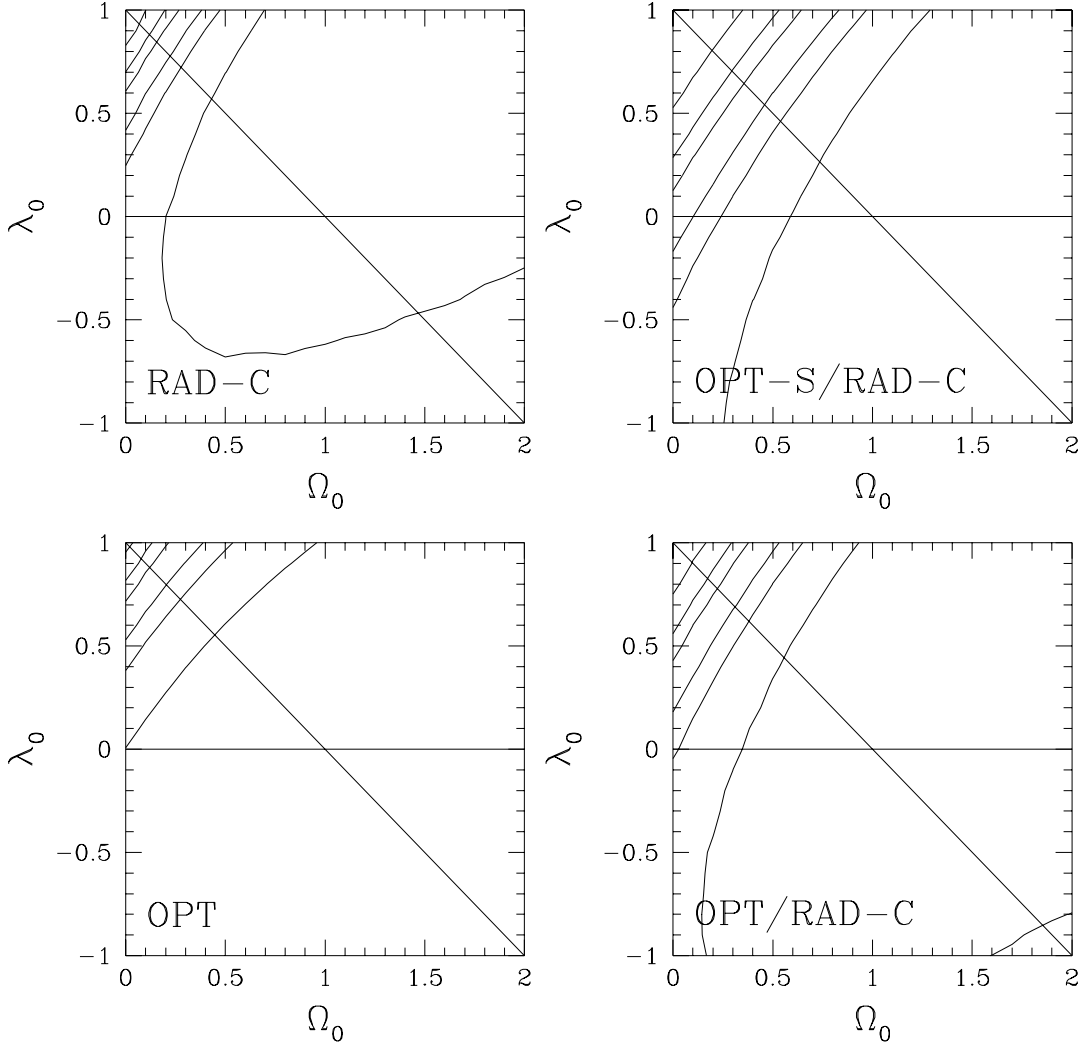}}
\figcaption{Likelihood contours in the $\Omega_0$-$\lambda_0$ plane.  The cases
  are radio only (top left), optical only (bottom left), joint optical
  and radio (bottom right), and joint optical and radio with spiral
  galaxies contributing in the optical (top right).  Contours are
  drawn at 68\%, 90\%, 95.4\%, 99\%, 99.7\%, and 99.99\% confidence
  intervals in the likelihood ratio for two degrees of freedom.  
  Note, however, that the maximum likelihood solution always lies
  on the edge of the $\Omega_0$--$\lambda_0$ grid.
  Flat models ($\Omega_0+\lambda_0=1$) lie along
  the diagonal line, and models with no cosmological constant
  ($\lambda_0=0$) lie along the horizontal line.  }
\label{f5}
\end{figure}

\begin{figure}
{\epsfxsize=15cm \epsfbox{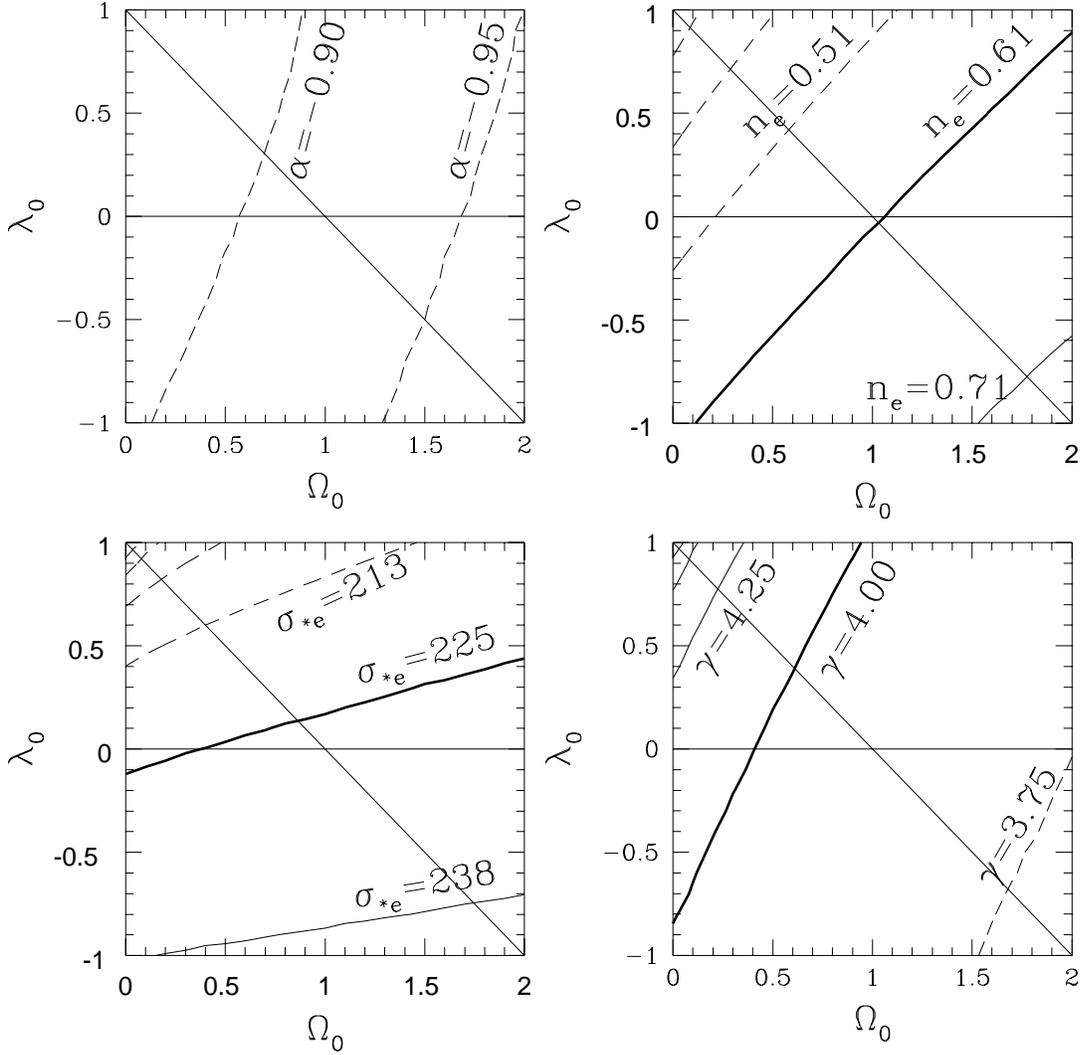}}
\figcaption{Luminosity function parameters as a function of the cosmological 
   model for model RAD-C-OPT. We show the Schechter $\alpha_e$ (top left), 
   comoving density $n_e$ ($10^{-2} h^3$ Mpc$^{-3}$, top right)
   dark matter velocity dispersion $\sigma_{*e}$ (km s$^{-1}$; bottom 
   left), and Faber-Jackson exponent $\gamma_e$ of the early-type
   galaxies (bottom right). The heavy solid line marks the best prior estimate,
   the solid (dashed) curves are spaced $0.5\sigma$ upwards (downwards)
   from the best estimate.  }
\label{f6}
\end{figure}

Tables \ref{models1} and \ref{models2} summarize the cosmological
results for our fits to both the radio and optical lens data for flat
($\Omega_0+\lambda_0=1$) and $\lambda_0=0$ cosmological models.  
The models are labeled ``RAD--A (B,C)"
which means the radio data with completeness model A (B,C), and ``OPT" or
``OPT--S" which means the optical data either without or with the
inclusion of a contribution from spiral galaxies.  The radio models
always include the spiral galaxies.  The best-fit cosmologies for the
two samples are statistically consistent, although the radio limits
are shifted to lower $\Omega_0$ by $\Delta \Omega_0 \simeq 0.1$.  For
the most conservative completeness Model C, the 2$-\sigma$ limit in a
flat cosmological model is $\Omega_0 > 0.27 $, compared to $\Omega_0 >
0.31$ for the optical data. A joint analysis of the optical and radio
for model C yields $\Omega_0 > 0.38$.  Changing to the radio
completeness models that bias the source distributions to higher
redshifts raises the limits by $\Delta\Omega_0 \simeq 0.05$.  We
generally do not obtain 2$-\sigma$ upper bounds on $\Omega_0$ over the
range $0 < \Omega_0 < 2$ because the lensing probability declines
slowly with higher $\Omega_0$ and because of the effects of Poisson
uncertainties for small numbers of objects.  Figure \ref{f5} shows
likelihood contours for the optical, radio, and joint analyses in the
$\Omega_0$-$\lambda_0$ plane for completeness model C, and Figure \ref{f6}
illustrates the shifts in the lens model parameters for the early-type 
galaxies as a function of the cosmological model.

\begin{figure}
{\epsfxsize=15cm \epsfbox{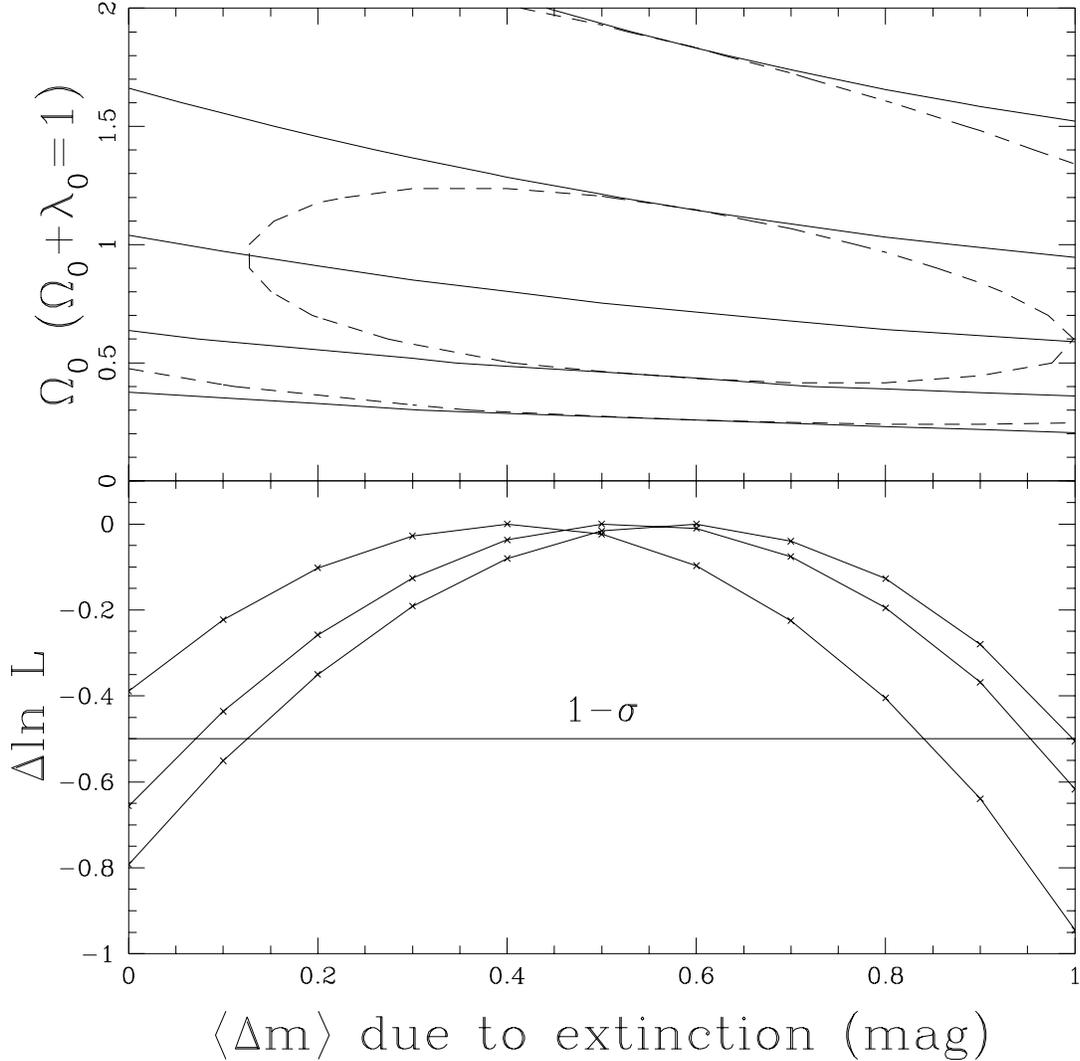}}
\figcaption{Extinction Estimates. The bottom panel shows the maximum 
  likelihood as a function of the mean magnitude change 
  $\langle\Delta m\rangle$
  produced by extinction for completeness models A (middle at left), 
  B (top at left), and C (bottom at left).  In the top panel we show
  likelihood contours as a function of $\Omega_0$ for flat cosmologies
  and completeness model C.  The solid lines show the maximum likelihood
  model and the 68\% and 95.4\% limits as a function of cosmology
  without the relative likelihoods for the $\dm$.  The dashed lines
  show the 68\% and 95.4\% limits on one parameter relative to the best fit 
  model including the relative likelihoods of the $\dm$. } 
\label{f7}
\end{figure}

\begin{figure}
{\epsfxsize=15cm \epsfbox{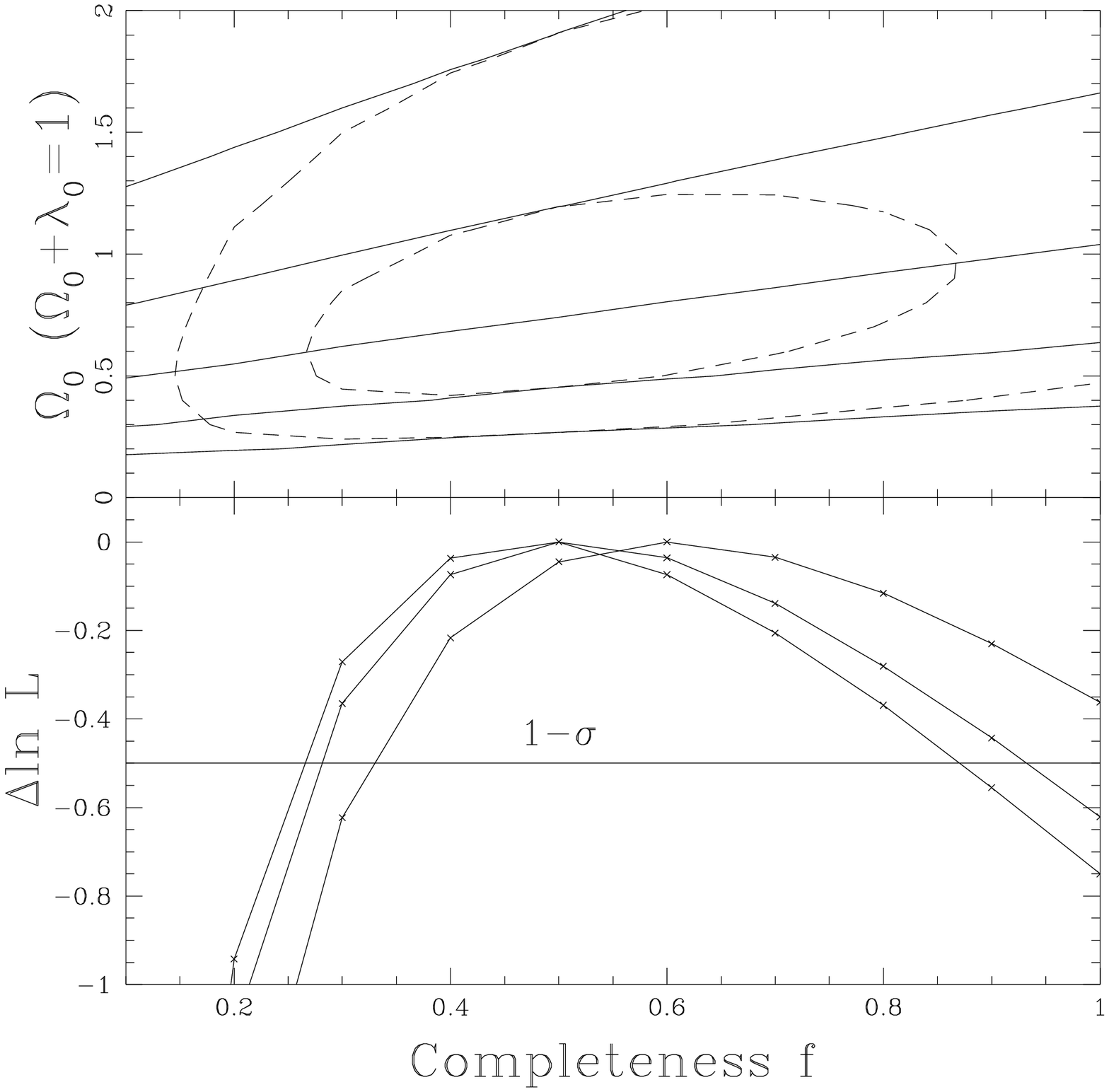}}
\figcaption{Completeness Estimates.  The bottom panel shows the maximum 
  likelihood as a function of the fractional completeness of the
  quasar lens sample $f$.  We can equivalently view $f$ as the
  fraction of early-type galaxies made opaque in the optical by
  extinction.  The top frame shows the limits on $\Omega_0$
  for flat cosmologies and completeness model C.  Notation is  
  the same as for Figure \ref{f7}. }
\label{f8}
\end{figure}

The addition of the constraint that the total comoving galaxy density
is far more certain than its division into spiral and early-type
galaxies increases the importance of the spirals in determining the
cosmological limits.  Previously, as we increased the cosmological
constant, the model would compensate by reducing the comoving
densities of both galaxy types.  Now, as the density of the early-type
galaxies decreases, the density of spirals increases because of the
constraint on the total galaxy density.  The extra optical depth of
the spirals limits the effectiveness of changes in the luminosity function 
in compensating for the change in cosmology and strengthens the
cosmological limits.  The quasars lensed by spirals are more affected
by both extinction (Kochanek 1996b; Perna, Loeb \& Bartelmann 1997)
and biases against including lensed quasars in quasar catalogs
(Kochanek 1991) than those lensed by 
early-type galaxies; therefore, we computed the optical
sample both with and without spiral galaxies.  We find better
consistency with the radio sample (at the 1$-\sigma$ level) if we
exclude spiral lenses from the optical sample; we use such a model as our
standard.

Even after excluding quasars lensed by spirals, 
the optical sample prefers cosmological models with lower optical depths than
the radio sample.  Equivalently, this sample may be more affected
by incompleteness due to extinction in the lens galaxies or systematic effects
of surveys for quasars on the lensed quasars.  To estimate the possible
level of extinction we computed the quasar lensing probabilities assuming
that each lens is intrinsically $\dm$ mag brighter than observed, where
$\dm$ is the mean magnitude change due to extinction.  The shift $\dm$
can also represent errors in the magnitude of the break in the quasar
luminosity function or offsets between the magnitude scale used for
the individual quasars and the quasar luminosity function.
Figure \ref{f7} shows the likelihood as
a function of $\dm$ and the corrected cosmological limits.  The best fit
value for the conservative completeness model C is $\dm = 0.58 \pm 0.45$, 
and zero extinction is ruled out at slightly better than 1$-\sigma$ in the 
likelihood ratio.  For the best fit model we still find $\Omega_0 > 0.26$ 
at 2$-\sigma$.  For completeness model A we find $\dm = 0.53\pm0.45$
and $\Omega_0 > 0.27 $, and for model B we find $\dm = 0.40 \pm 0.45$ and
$\Omega_0 > 0.33$.  A negative $\Delta m$ is possible; it corresponds
either to an error in the quasar LF break magnitude or incompleteness
in the radio sample.  For the Seaton (1979) model of the UV extinction
curve and a typical lens redshift of $0.5$, $\dm \simeq 6 E(B-V)$; thus, 
the magnitude changes correspond to $E(B-V) \simeq 0.1$.
We can also give the quasars a relative completeness of $f$ compared to
the radio sources by rescaling the comoving density of galaxies lensing
the quasars to a fraction $f$ of the true density.  The model mimics both
incompleteness of the lensed quasar samples due to biases in quasar
searches against lensed quasars (possibly a 10--20\% effect, Kochanek 1991),
and a bimodal extinction distribution for galaxies with fraction $f$
transparent and fraction $1-f$ opaque.  As illustrated in Figure \ref{f8},
the best fit for completeness model C is $f=0.48$ with a 1$-\sigma$
range from $0.26 < f < 0.87$ and a 2$-\sigma$ bound of $\Omega_0 >0.25$
in flat models. For Model A the best fit is $f=0.52$ ($0.28 < f < 0.93$)
and for Model B the best fit is $f=0.61$ ($0.33 < f \lesssim 1.0$).  
In both models there is evidence that the quasar lens sample is
significantly incomplete, although the significance of the result
is weak because of the large Poisson uncertainties in the two samples.

\section{Conclusions}

We are conducting a redshift survey of 177 flat-spectrum radio sources in
the flux range from 50 to 250 mJy; we have measured 124 redshifts that enabled
us to estimate the cosmological model from the statistics of the 6
lensed radio sources found in the JVAS survey 
(Patnaik \etal\ 1992a; Patnaik 1994; 
King \etal\ 1996; Browne \etal\ 1997) of 2500
flat-spectrum sources brighter than 200 mJy.  The mean redshifts of
the sources are $\langle z \rangle \simeq 1.2$; they show little
variation with the radio flux below 500 mJy.  A rapidly increasing
fraction of the sources consists of radio galaxies rather than radio
quasars, rising from $\sim$ 10\% of sources brighter than 1 Jy to
$\sim$ 50\% of sources at 100 mJy.  The rapid change in the source
population from blue quasars to red galaxies means that radio lenses
will typically be redder than radio sources of the same flux and that
many lensed flat-spectrum radio sources will
show extended lensed optical emission, as already observed in MG~0414+0534 
(\cite{aw94}; \cite{la95}; \cite{fa97}) and CLASS~1608+656 (\cite{class95};  
Jackson \etal\ 1997).

The cosmological limits from the radio-selected sample are
statistically consistent with those derived from lensed quasars (see
Maoz \& Rix 1993, Kochanek 1993, 1996a).  The 2$-\sigma$ limits
on the cosmological constant in flat models for the radio, optical,
and joint samples using the most conservative model for the
completeness of our redshift surveys and excluding quasars lensed
by spirals are $\Omega_0 > 0.27$, $0.31$,
and $0.38$ respectively.  The 1$-\sigma$ limits are $0.47 < \Omega_0 < 1.38$, 
$0.65 < \Omega_0$, and $0.64 < \Omega_0 < 1.66$. The small
numbers of lenses and the slow decline of the optical depth for large
$\Omega_0$ mean that we find no 2$-\sigma$ upper bound for all
three models and no 1$-\sigma$ upper bound for the optical sample up
to the limit of $\Omega_0 < 2$.  The weaker variation of the optical
depth for cosmologies without a cosmological constant gives us only
1$-\sigma$ lower bounds on $\Omega_0$ of 0.24, 0.30, and 0.51
respectively.  For comparison, Perlmutter \etal\ (1997) formally obtained 
1$-\sigma$ limits of $0.66 < \Omega_0 < 1.28$ for flat models and
$0.28 < \Omega_0 < 1.57$ for $\lambda_0=0$ models using Type Ia
supernovae.  Their formal statistical limits did not include several
of the expected systematic uncertainties 
(\eg, extinction, K-correction models, 
Malmquist biases) which can produce additional uncertainties of 
$\Delta \Omega_0 \sim 0.2$ (Perlmutter \etal\ 1997).  
The essentially perfect statistical consistency of these two radically
different methods and the internal consistency of the radio and optical
lens methods for determining the global cosmological model is a
very reassuring check that neither approach has missed a catastrophic
systematic error.

The agreement between the optical and radio samples is better (by
about 1$-\sigma$) if there is no spiral galaxy contribution to the
lensed quasars; we adopted a spiral-free model 
as our standard.  The
optically selected samples are biased against the inclusion of spirals
both because of the higher extinction expected for spirals compared to
early-types and because optical surveys to find quasars are more
likely to exclude lenses produced by spirals than those produced by
early-types.  Kochanek (1991) noted that color and spectral selection
methods for finding quasars were intrinsically biased against
including lensed quasars, but that for bright quasars ($m <19$) lensed
by early-type galaxies the effects were small ($\lesssim 10$\%).
However, a spiral galaxy lens is 1 to 2 mag brighter than an
early-type lens for the same image separation and more easily masks
the presence of the lensed quasar images.  Where we have optical
images of the quasar lens galaxies, they are all consistent with
early-type galaxies, while a few of the radio lenses appear to be
spirals (see Keeton, Kochanek \& Falco 1997 for
a summary).  It is thus plausible
that the combined effects of extinction and the optical quasar survey
biases have eliminated most spiral lenses from the quasar lens sample.

Even after eliminating the spiral galaxies from the quasar analysis,
the best fit optical depth implied by the quasars is less than that
implied by the radio sources.  We may be overestimating the radio
optical depth by including the two lenses with significant extended
radio structure (B~1938+666 and B~2114+022), because extended sources
can have significantly higher probabilities of being lensed (Kochanek
\& Lawrence 1990).  Although only 5\% of the JVAS sources are extended
or multi-component (Patnaik \etal\ 1992b), $15-30$\% of the lensed
sources are extended or multi-component depending on whether
B~2114+022 is 2 images of a double source or 4 images of a single
source (see Browne \etal\ 1997). Alternatively, we may be
underestimating the quasar optical depth by neglecting either
extinction in early-type galaxies (Kochanek 1991, 1996a; Tomita 1995;
\cite{mal96}) or the biases in quasar surveys against lensed
quasars (Kochanek 1991).  When we fitted models where the mean lensed
quasar is $\dm$ mag fainter due to extinction, we found $\dm = 0.58
\pm 0.45$; when we fitted models in which a fraction $f$ of quasar
lenses are lost due to survey biases or opaque galaxies we found $f =
0.48_{-0.22}^{+0.39}$. In both cases the model with no selection
effect is ruled out at slightly over 1$-\sigma$.  Even so, the limits
on the cosmological constant remain reasonably robust, with $\Omega_0
> 0.26$ at 2$-\sigma$ in flat models.  A mean magnitude change of $\dm
\sim 0.6$ corresponds to $E(B-V) \simeq 0.1$ in the lens galaxy, a
$B-V$ color change of about $0.1$ mag and a $B-K$ color change of
approximately $0.5$ mag.

Malhotra \etal\ (1996) have argued for a far larger effect from dust
on the optically selected samples, largely by comparing the colors of
radio selected lenses to the colors of optically selected lenses.
They advocate a mean magnitude change of $\dm \sim 2 \pm 1$, or about
4 times our estimate.  We believe Malhotra \etal\ (1996) were
comparing the colors of intrinsically different populations, leading
them to overestimate the effects of extinction.  We see the population
shift in our redshift survey with the rapidly rising fraction of
(early-type) galaxies at fainter radio fluxes, and we see the
population shift in the lensed sources by the frequent appearance of
extended lensed optical emission (MG~0414+0534, CLASS~1608+656) and
the frequent lack of the broad emission lines characteristic of
quasars.  Most of the observed lenses where there is a
consensus for extinction of the source by the lens galaxy are clearly
spiral lenses either because we directly observe the spiral structure
(2237+0305, Nadeau \etal\ 1991) or from the presence of atomic and
molecular gas (B~0218+357 and PKS~1830--210, Carilli, Rupen, \&
Yanny 1993; Wicklind \& Combes 1996; Lovell \etal\ 1996).  The
exception is MG~0414+0534 (\cite{la95}), where the lens
galaxy itself is far redder than any passively evolving early-type
galaxy or any other lens galaxy (Falco \etal\ 1997; Keeton,
Kochanek \& Falco 1997).  We also cannot use MG~0414+0534 to make
simple inferences about lensed quasars because the very presence of
extended lensed optical emission means that the source is very
different from a bright optical quasar.  A likely counter hypothesis
to that of Malhotra \etal\ (1996) is that all extremely red lensed sources
will turn out to have extended lensed optical emission in deep HST
images.

The general agreement of the radio and optical samples does not
eliminate the question of common systematic dependencies.  The most
important problem is the continuing uncertainty in the galaxy
luminosity function, particularly when divided by type; the
statistical uncertainties in the present models are roughly equally
due to the Poisson uncertainties from the numbers of lenses and the
uncertainties in the local comoving density of early-type galaxies.
The models do not include any evolution in the galaxy populations,
although models of lensing with galaxy evolution (Mao 1991; Mao \&
Kochanek 1994; Rix \etal\ 1994) demonstrated that lens statistics are
considerably less sensitive to galaxy evolution than one might naively
expect. Most plausible merger models conserve the optical depth while
changing the separation distribution.  Moreover, since the mean lens
redshift is usually less than $z=1$, redshift surveys have
already confirmed that the early-type population that dominates
gravitational lensing shows little evolution (e.g. Lilly et al. 1995).  Nonetheless, evolution
is a significant systematic question that we should address
in greater detail.

Our ability to expand the cosmological conclusions is largely
restricted by the need for additional redshift data.  The incompleteness
of our redshift survey leads to uncertainties of $\Delta\Omega_0\simeq0.1$
in the cosmological model, and despite our
survey, we still cannot include the majority of the known
radio lenses found in systematic surveys in our analysis for lack of data
on the luminosity function.  Interpreting the 
CLASS survey (5 lenses so far) requires that the flat-spectrum
redshift distribution be extended to $\sim 5$ mJy, and interpreting
the MG Survey (5 lenses so far) requires that the steep-spectrum redshift
distribution be extended to $\sim 100$ mJy.  The lensing optical depth
varies strongly along lines of constant $\Omega_0+\lambda_0$, and
rather weakly in the orthogonal direction, leading to the 
degenerate likelihood contours in the $\Omega_0$-$\lambda_0$ plane
shown in Figure \ref{f5}.  One way to break the degeneracy, and also to
strengthen the overall limits, is to use the distribution of lens
galaxy redshifts compared to source redshifts (Kochanek 1992), because 
the mean lens redshift has a different dependence on $\Omega_0$ and $\lambda_0$
than the optical depth (Kochanek 1993).  Unfortunately, both the
source and lens redshifts are known for only a small fraction of the 
lenses, and the corrections for incompleteness when using
the lens redshifts are both important and difficult to model (\cite{hk96}; 
Kochanek 1996a).

\acknowledgments  
We particularly 
thank the MG Collaboration, J. Hewitt, B. Burke, L. 
Herrold, and J. Leh\'ar for supplying us with samples of
flat-spectrum sources.  We also thank P. Schechter for
MDM 2.4m images, L. Macri for MMT spectra and 
P. Berlind for Tillinghast spectra of a number of our sources.
CSK is supported by NSF grant AST-9401722 and
NASA ATP Grant NAG5-4062. JAM is supported by a postdoctoral
fellowship from the Ministerio de Educaci\'on y Cultura, Spain.

\clearpage


\clearpage
\begin{deluxetable}{lccccllll}
\scriptsize
\tablecaption{Sample 1 (JVAS  200--250 mJy)}
\tablehead{
Object  &  $\alpha$ (B1950)&    $\delta$ (B1950)  & $ I $ &  $\sigma_I$  &  \multicolumn{1}{c}{$z$} & \multicolumn{1}{c}{$\sigma_z$} & 
\multicolumn{2}{l}{type \hspace{1cm} detected lines}}
\startdata
0902+468  &09 02 52.68 & 46 48 21.71 & 14.8  & 0.2  &  0.0848  &  0.0005 & E &  (HK, H$\delta$, G, Mg, CaFe, Na), H$\alpha$, OIII  \nl
0903+669   &09 03 01.85 & 66 56 51.58 & 18.9 & 0.2 & &  & \nl
0905+420   &09 05 20.99 & 42 02 56.14 & 18.2 & 0.1 &     0.7325  &  0.0008 &  Q   &  \ciii, MgII, H$\gamma$, H$\beta$  \nl
0920+416   &09 20 19.92 & 41 38 20.60 & 18.0 & 0.1 &     0.028  &  0.001  & L & (HK, H$\beta$, Mg, CaFe, Na), H$\alpha$, OIII \nl
0924+732   &09 24 51.83 & 73 17 12.42 & 18.6 & 0.1 & & & D & \nl
0927+469   &09 27 17.71 & 46 57 20.96 & 16.8 & 0.2 &     2.032   &  0.001  &  Q   &  Ly$\alpha$, SiIV, CIV, \ciii\   \nl
0927+586   &09 27 10.76 & 58 36 35.55 & 17.1 & 0.1 &     1.9645  &  0.0009 &  Q   &  Ly$\alpha$, SiIV, CIV, HeII, \ciii\  \nl
0939+620   &09 39 29.44 & 62 04 17.76 & 18.0 & 0.2 &     0.7533  &  0.0005 &  Q   &  MgII, NeV  \nl
0951+422   &09 51 06.97 & 42 15 20.74 & 19.3 & 0.1 &     1.783   &  0.004  &  Q   &  SiIV,  CIV, \ciii, MgII \nl
0955+509   &09 55 22.22 & 50 54 18.83 & 17.7 & 0.2 &     1.154   &  0.002  &  Q   &  CIV, \ciii, CII, MgII, HeI  \nl
1010+495   &10 10 20.75 & 49 33 33.83 & 18.5 & 0.1 &     2.201   &  0.002  &  Q   &  Ly$\alpha$, CIV, \ciii\  \nl
1023+747   &10 23 13.02 & 74 43 44.02 & 17.5 & 0.2 &    0.879  &  0.002  & Q  &  MgII, OIII, OIV  \nl
1027+749$^\dagger$ &10 27 13.30 & 74 57 23.11 & 15.2 &0.2     &  0.123    & 0.001        & E &                         \nl 
1028+564   &10 28 50.61 & 56 26 23.42 & 21.5 & 0.5 &        &  & D & \nl
1101+609   &11 01 50.75 & 60 55 07.10 & 18.9 & 0.2 &     1.363   &  0.003  &  Q   &  CIV, \ciii, MgII  \nl
1109+350   &11 09 55.21 & 35 02 58.82 & 19.1 & 0.2 &     1.9495  &  0.0003 &  Q   &  Ly$\alpha$, CIV, \ciii\  \nl
1116+603$^\dagger$ &11 16 19.23 & 60 21 22.49 & 17.5 &0.2     &  2.638    &         &  Q   &                      \nl
1117+543   &11 17 33.00 & 54 20 53.33 & 18.8 & 0.2 &     0.924   &  0.001  &  Q   &  \ciii, MgII, OIIIa, NeV, H$\gamma$ \nl
1131+730   &11 31 11.77 & 73 05 55.21 & 18.2 & 0.2 &     1.571   &  0.002  &  Q   &  SiIV, CIV, HeII, \ciii, MgII  \nl
1147+438   &11 47 39.81 & 43 48 47.00 & 18.8 & 0.1 &     3.037   &  0.008  &  N  &  Ly$\alpha$, CIV, \ciii\ \nl
1151+598   &11 51 24.00 & 59 51 35.93 & 19.9 & 0.2 &     0.871   &  0.002  &  Q   &  \ciii, MgII, H$\gamma$  \nl
1200+468   &12 00 58.77 & 46 49 37.77 & 21.4 & 0.2 & &  & \nl
1200+608   &12 00 30.71 & 60 48 01.36 & 14.4 & 0.1 &     0.0656  &  0.0002 & E &  HK, H$\delta$, Mg, CaFe, Na  \nl
1204+399   &12 04 04.63 & 39 57 45.72 & 18.2 & 0.2 &     1.5134  &  0.0009 &  Q   &  CIV, \ciii, CII, MgII  \nl
1231+507   &12 31 27.08 & 50 42 54.89 & 16.7 & 0.1 &     0.2075  &  0.0005 & E &  HK, G, Mg, CaFe Na  \nl
1234+396   &12 34 26.25 & 39 36 57.85 & 19.0 & 0.2 & & & D & \nl
1238+702   &12 38 32.70 & 70 14 57.98 & 16.6 & 0.1 &     1.4706  &  0.0005 &  Q   &  CIV, \ciii, MgII  \nl
1239+606   &12 39 16.55 & 60 37 08.06 & 16.8 & 0.1 &     1.457   &  0.005  &  N  &  SiIV, CIV, NIII, \ciii\ \nl
1245+676   &12 45 32.18 & 67 39 38.12 & 16.9 & 0.2 &              0.1073  &  0.0002 & E & (HK, H$\delta$, G, H$\beta$, Mg, CaFe, Na)  \nl
1245+716   &12 45 15.69 & 71 40 41.97 & 20.8 & 0.3 &          & & D  & \nl
1300+485   &13 00 03.36 & 48 35 24.34 & 16.1 & 0.2 &     0.873   &  0.001  &  Q   &  \ciii, CII, MgII, HeI  \nl
1300+693   &13 00 50.97 & 69 18 57.72 & 17.1 & 0.2 &     0.5677  &  0.0003 & L &  CII, NeV, OII, HeI, H$\gamma$, OIII  \nl
1302+356   &13 02 15.38 & 35 39 57.94 & 21.7 & 0.3 & &  & \nl
1310+487   &13 10 32.94 & 48 44 24.63 & 19.3 & 0.2 & (0.313)  &  0.003  & L &  OIII, NeV,  H$\gamma$, H$\beta$  \nl
1318+508   &13 18 36.32 & 50 51 50.13 & 21.2 & 0.7 &         & &  D & \nl
1327+504   &13 27 02.23 & 50 24 55.57 & 18.1 & 0.2 &     2.654   &  0.001  &  Q   &  OVI, SIV, Ly$\alpha$, SiIV, CIV \nl
1328+523   &13 28 41.69 & 52 17 41.92 & 19.3 & 0.2 &	        & & D  & \nl
1339+696   &13 39 29.98 & 69 38 30.80 & 18.7 & 0.2 &     2.255   &  0.003  &  B  &  Ly$\alpha$, CIV, \ciii\ \nl
1341+691   &13 41 42.19 & 69 10 21.11 & 17.3 & 0.2 &     1.417   &  0.002  &  Q   &  CIV, \ciii, CII, MgII \nl
1349+618   &13 49 01.61 & 61 47 37.87 & 20.7 & 0.4 &	        1.834  &  0.002  &  Q   &  CIV, NIII, \ciii, NeIV  \nl   
1409+595   &14 09 49.22 & 59 31 08.20 & 20.1 & 0.2 &     1.725   &  0.009  &  Q   &  CIV, \ciii, MgII  \nl
1412+461   &14 12 19.18 & 46 08 46.22 & 19.9 & 0.2 &    0.186  &  0.002  & E &  (HK, H$\delta$, CaFe, Na)  \nl
1418+375   &14 17 55.81 & 37 35 18.25 & 17.9 & 0.1 &     0.969   &  0.002  &  B&  NIII, \ciii, MgII \nl
1419+469   &14 19 30.38 & 46 59 27.87 & 16.2 & 0.2 &     1.665   &  0.003  &  Q   &  SiIV, CIV, \ciii, MgII \nl
1421+511$^\dagger$ & 14 21 28.55 & 51 09 12.34 & 15.0 &0.2     &  0.274    &  0.002       &  Q   &                             \nl 
1427+634   &14 27 52.03 & 63 29 23.84 & 20.9 & 0.2 &     1.561   &  0.001  &  Q   &  CIV, HeII, \ciii, CII, MgII  \nl
1438+501   &14 38 04.29 & 50 10 56.24 & 17.7 & 0.2 & 0.174  &  0.002  & E &  (HK,  H$\delta$, G, Mg, H$\beta$, CaFe)  \nl
1447+536   &14 47 26.02 & 53 38 33.49 & 22.1 & 0.5 &	        &  & \nl
1450+455$^\dagger$ & 14 50 37.18 & 45 34 38.12 & 16.0 & 0.2     &  0.469    &         & E &  \nl  
1454+447   &14 54 06.02 & 44 43 41.66 & 17.8 & 0.2 & &  & \nl
1533+487   &15 33 42.16 & 48 46 54.20 & 16.2 & 0.2 &     2.563   &  0.002  &  N  &  Ly$\alpha$, CIV, \ciii\ \nl
1556+745   &15 56 54.94 & 74 29 32.56 & 19.3 & 0.2 &     1.667   &  0.001  &  Q   &  CIV, HeII, \ciii, MgII \nl
1557+565   &15 57 41.57 & 56 33 41.87 & 16.0 & 0.1 &	        0.30 &  0.03 & E &  (HK, H$\delta$, G)  \nl
1558+595   &15 58 05.76 & 59 32 48.42 & 15.0 & 0.2 &	            0.0602  &  0.0001 & E &  (HK, H$\delta$, G, H$\beta$, Mg, CaFe, Na)  \nl
1603+573   &16 03 34.72 & 57 22 42.20 & 16.3 & 0.2 &     0.720   &  0.001  &  Q   &  \ciii, MgII, NeV, H$\gamma$, H$\beta$  \nl
1611+425   &16 11 25.57 & 42 30 52.93 & 20.3 & 0.4 &	        &  & \nl
1627+476   &16 27 11.18 & 47 40 42.41 & 18.4 & 0.2 &     1.629   &  0.001  &  Q   &  SiIV, CIV, HeII, \ciii, CII, MgII \nl
1646+411   &16 46 50.96 & 41 09 16.65 & 20.0 & 0.2 &	            0.8508  &  0.0003 &  Q   &  \ciii, MgII, H$\gamma$  \nl
1646+499   &16 46 16.48 & 49 55 14.75 & 14.1 & 0.2 &     0.0475  &  0.0001 & L &  (HK, G, Mg, CaFe, Na), H$\alpha$, OIII  \nl
1650+581   &16 50 31.80 & 58 10 39.84 & 22.5 & 1.0 &	        &  & \nl
1655+534   &16 55 32.40 & 53 26 24.60 & 16.9 & 0.2 &     1.553   &  0.002  &  Q   &  CIV, \ciii, MgII  \nl
1704+512   &17 04 13.38 & 51 13 34.34 & 16.7 & 0.2 &     0.5303  &  0.0003 &  Q   &  MgII, NeV, HeI, OIII  \nl 
1712+493   &17 12 17.48 & 49 19 56.91 & 19.3 & 0.2 &     1.552   &  0.002  &  Q   &  CIV, HeII, \ciii, MgII \nl
1738+451   &17 38 39.49 & 45 08 20.42 & 15.7 & 0.2 &     2.788   &  0.008  &  N  &  Ly$\alpha$, CIV, \ciii\ \nl
1742+378   &17 42 05.62 & 37 49 08.35 & 16.4 & 0.2 &     1.9578  &  0.0005 &  Q   &  Ly$\alpha$, SiIV, CIV, HeII, \ciii\  \nl
1745+643$^\dagger$  &17 45 51.98 & 64 22 50.89 & 20.8 & 0.3 	          &  1.228    &         & E & \nl
1750+509   &17 50 21.11 & 50 56 17.43 & 16.5 & 0.2 &     0.3284  &  0.0004 & L &  (HK, H$\delta$, G, Mg, CaFe), Mg, OII, H$\gamma$, OIII  \nl
1752+356   &17 52 27.92 & 35 41 17.64 & 16.8 & 0.2 &     2.207   &  0.002  &  Q   &  Ly$\alpha$, CIV, \ciii\ \nl
1755+626   &17 55 23.68 & 62 37 03.36 & 15.1 & 0.4 & 	          0.0276  &  0.0001 & E &  (HK, H$\beta$, Mg, CaFe, Na)  \nl
\hline
\enddata
\small 
\tablecomments{
A $\dagger$ indicates a previously known source as per NED, for which we
did not obtain spectra; HK and G are the CaII H\&K lines and G bands,
respectively; parentheses surrounding a list of lines indicate
absorption; parentheses surrounding a redshift indicate a marginal
measurement; D, E, L, Q, B, N and b indicate respectively a detected
object, an early-type galaxy, a late-type galaxy, a quasar, a quasar
with broad absorption lines, a quasar with narrow absorption lines and
a BL Lac object.
}
\label{s1id}
\end{deluxetable}

\clearpage

\begin{deluxetable}{lccccllll}
\scriptsize
\tablecaption{Sample 2 (MG  100--200 mJy)}
\tablehead{
\colhead{Object}  &  
\colhead{$\alpha$ (B1950)} &    
\colhead{$\delta$ (B1950)} & 
\colhead{$I$} & \colhead{$\sigma_I$} & \colhead{$z$} & \colhead{$\sigma_z$} & 
\multicolumn{2}{l}{\mbox{type \hspace{1cm} detected lines}}}
\startdata
MGC0001+2113   & 23 58 58.58 & 56 54.04 & 17.7 & 0.1 &	  1.106   &  0.002  &  Q   &  \ciii, NeV, HeI  \nl
MGC0034+3712   & 00 32 14.32 & 55 53.66 & 18.9 & 0.2  &	  1.390   &  0.002  &  Q   &  CIV, \ciii, MgII  \nl
MGC0037+2613   & 00 34 40.35 & 56 43.50  &      &   &	          0.1477  &  0.0002 & E &  (HK, G, H$\beta$, Mg)   \nl
MGC0042+2739   & 00 39 55.71 & 23 22.41  &      &   &	         & & & \nl
MGC0046+2249   & 00 43 41.10 & 33 20.37  &      &   &	         & & & \nl
MGC0046+2456   & 00 43 28.10 & 40 09.40 & 17.1 & 0.2 &	  0.7467  &  0.0004 &  Q   &  NeIV, MgII, HeI  \nl
MGC0054+2549   & 00 51 54.96 & 33 49.06  &            &        	         & & & & \nl
MGC0054+3842   & 00 51 27.85 & 25 58.52  &            &        	         & & & & \nl
MGB1606+2031   & 16 03 54.30 & 40 12.40 &  &  &	         & & & \nl
MGB1634+1946   & 16 32 34.50 & 53 14.76 & 17.9 & 0.2  &	  0.792   &  0.003  &  Q   &  \ciii, CII, MgII, HeI  \nl
MGB1655+1949   & 16 53 32.99 & 53 29.07 & 16.6 & 0.2  &	  3.260   &  0.003  &  N  &  Ly$\beta$, Ly$\alpha$, SiIV, CIV \nl
MGB1705+2215   & 17 03 22.21 & 20 08.25  &            &   &	  0.04977 &  0.00008 & E &  (HK, G, H$\beta$, Mg, CaFe, Na)  \nl
MGB1715+3619   & 17 13 22.85 & 23 08.90 & 18.4  & 0.2  &	  0.5549  &  0.0003 &  Q   &  MgII, HeI, H$\beta$, OIII  \nl
MGB1720+2334   & 17 18 05.64 & 38 29.12 & 17.4  & 0.2  &	  1.852   &  0.003  &  Q   &  Ly$\alpha$, SiIV, CIV, \ciii, MgII  \nl
MGB1728+1931   & 17 26 44.62 & 33 31.38 &   &  &	          0.1756  &  0.0003 & E &  (HK, G, H$\beta$, Mg, CaFe, Na)  \nl
MGB1745+2252   & 17 42 59.09 & 53 57.86 & 17.5  & 0.2 &	  1.8838  &  0.0007 &  Q  &  Ly$\alpha$, SiIV, CIV, HeII, \ciii\ \nl
MGB1747+2323   & 17 45 45.21 & 25 37.51 & 17.1  & 0.1 &	  2.203   &  0.002  &  N  &  Ly$\beta$, Ly$\alpha$, SiIV, CIV  \nl
MGB1807+3107   & 18 05 38.33 & 05 52.75 & 18.3  & 0.2 &	  0.5373  &  0.0004 &  N  &  MgII, NeV, OIII  \nl
MGB1813+3144$^\dagger$& 18 11 42.73 & 43 22.31& 16.3  & 0.1  	 &  0.117   &         & b &   \nl
MGB1834+2051   & 18 32 03.59 & 49 16.53 & 16.8  & 0.2 &	 & &D & \nl
MGB1835+2506   & 18 33 55.57 & 04 13.20 &   &  &	          1.9728  &  0.0009 &  B  &  CIV, \ciii\ \nl
MGB1843+3150   & 18 41 10.08 & 47 23.59 & 15.9 & 0.1&	  0.4477  &  0.0003 &  Q   &  MgII, NeV, HeI, H$\gamma$, H$\beta$  \nl
MGB1843+3225   & 18 41 37.21 & 22 22.47 & 16.8 & 0.3  &	 & &D & \nl
MGB1846+2036   & 18 43 55.22 & 32 54.81 & 16.9 & 0.1      &	 & &D &    \nl
MGB1853+2344   & 18 51 22.48 & 40 48.28 & 14.1 & 0.2  &	  1.0311  &  0.0008 &  Q   &  \ciii, MgII  \nl
MGC2036+2227   & 20 34 44.58 & 17 29.07 & 16.4 & 0.1 &	  2.567   &  0.002  &  Q  &  Ly$\alpha$, SiIV, CIV, \ciii\ \nl
MGB2043+2256   & 20 41 40.27 & 46 26.50 & 16.7 & 0.1 &	  1.0810  &  0.0003 &  Q   &  \ciii, MgII  \nl
MGB2051+1950   & 20 48 56.61 & 38 48.99 & 16.6 & 0.1 &	  2.365   &  0.002  &  Q   &  Ly$\alpha$, SiIV, CIV, \ciii\  \nl
MGC2054+2407   & 20 52 17.47 & 56 05.77 & 16.5 & 0.2  &	  1.3774  &  0.0005 &  Q   &  CIV, \ciii, MgII  \nl
MGC2100+2058   & 20 57 49.45 & 47 34.81 & 17.6 & 0.1 &  (0.19)  &  0.04   & E &  HK \nl
MGC2100+2346   & 20 57 51.93 & 35 17.88 & 17.0 & 0.2 &	  1.124   &  0.001  &  Q   &  CIV, HeII, \ciii, MgII  \nl
MGC2100+2615   & 20 58 28.63 & 03 49.70 &  &  &	         & &D & \nl
MGC2105+2920   & 21 03 35.78 & 08 49.82 & 18.6 & 0.1  &	  1.347   &  0.002  &  Q   &  CIV, \ciii, MgII  \nl
MGC2106+2135   & 21 03 55.28 & 23 31.85 & 17.8 & 0.1  &	  0.6469  &  0.0008 &  Q   &  MgII, OII, OIII  \nl
MGC2109+2154   & 21 06 53.16 & 42 50.03 & 17.6 & 0.1 &	  2.344   &  0.002  &  N  &  Ly$\alpha$, CIV, \ciii\ \nl
MGC2109+2211   & 21 07 40.17 & 00 00.30 & 18.4 & 0.1 &	  2.281   &  0.002  &  Q   &  Ly$\alpha$, CIV, \ciii\ \nl
MGC2116+3016   & 21 13 59.43 & 04 05.38 & 17.3  & 0.2 &	  2.080   &  0.003  &  N  &  Ly$\alpha$, SiIV, CIV, HeII, \ciii\ \nl
MGC2118+2006   & 21 16 08.43 & 54 54.12 &   &  & 	 & & &    \nl
MGC2125+2441   & 21 23 11.64 & 29 00.28 &   &  &	         & & & \nl
MGC2130+3332   & 21 28 22.92 & 19 35.42 & 17.9  & 0.2  &	  1.473   &  0.006  &  Q   &  CIV, \ciii, MgII  \nl
MGC2137+2357   & 21 34 49.60 & 43 31.15 & 17.1  & 0.2  &	  0.6044  &  0.0007 &  Q   &  MgII, NeV, HeI, H$\gamma$, H$\beta$, OIII  \nl
MGC2153+2351   & 21 50 45.69 & 37 48.69 &   &  &	         & & & \nl
MGC2203+3712   & 22 01 08.58 & 56 45.03 & 14.5  & 0.2 &	  1.817   &  0.005  &  N  &  Ly$\alpha$, CIV, MgII  \nl
MGC2213+2558   & 22 11 27.17 & 43 30.11 & 15.0  & 0.3  &	  0.0940  &  0.0002 & E &  (HK, H$\delta$, G, H$\beta$, Mg, CaFe, Na)  \nl
MGC2214+3550   & 22 12 44.73 & 36 29.15 & 18.2  & 0.2 &	 0.877   &  0.003  &  Q   &  \ciii, CII, MgII  \nl
MGC2214+3739   & 22 11 55.07 & 24 14.24 &   &  & 	 & & &    \nl
MGC2223+2439   & 22 20 47.66 & 24 00.50 & 17.7  & 0.1 &	  1.490   &  0.004  &  Q   &  SiIV, CIV, HeII, \ciii, CII, MgII  \nl
MGC2227+3716   & 22 25 04.20 & 59 59.08 & 17.4  & 0.2 &	  0.975   &  0.003  &  N  &  HeII, \ciii, MgII, H$\gamma$  \nl
MGC2229+3057   & 22 27 15.93 & 41 48.78 &   &  &	          0.3196  &  0.0004 & L &  NeV, HeI, H$\gamma$, H$\beta$, OIII  \nl
MGC2230+2752   & 22 27 55.30 & 38 18.77 &   &  &	         & & & \nl
MGC2250+3825   & 22 47 48.11 & 08 42.70 &   &  & 	          0.1187  &  0.0003 & E &  (HK, G, H$\beta$, Mg, CaFe, Na)  \nl
MGC2251+2217   & 22 49 27.88 & 01 40.50 & 20.2  & 0.2  &	  3.668   &  0.003  &  N  &  SIV, Ly$\alpha$, CII, CIV  \nl
MGC2254+2058   & 22 52 27.05 & 42 36.60 &   &  &	          0.0635  &  0.0002 & E &  (HK, H$\delta$, G, H$\beta$, Mg, CaFe, Na)  \nl
MGC2257+3706   & 22 55 15.49 & 50 26.43 &   &  &	         & & & \nl
MGC2301+3512   & 22 58 52.85 & 56 52.64 &   &  &	          0.1357  &  0.0005 & E &  (HK, G, Mg, CaFe, Na), H$\alpha$, OIII  \nl
MGC2308+2008   & 23 05 43.49 & 52 26.67 &   &  &	          0.2342  &  0.0007 & L &  MgII, H$\beta$, OIII, H$\alpha$  \nl
MGC2309+3726   & 23 06 51.15 & 09 53.28 & 18.3  & 0.2  &	 & &D & \nl
MGC2315+3727   & 23 12 44.93 & 10 32.86 &   &  &	         & & & \nl
MGC2318+2404   & 23 16 05.73 & 48 14.79 & 17.2  & 0.1  &	 & &D & \nl
MGC2344+3433   & 23 42 20.80 & 17 09.05 & 17.8  & 0.2  &	  3.053   &  0.007  &  B  &  SVI, Ly$\beta$, Ly$\alpha$, CIV, \ciii\ \nl 
MGC2348+3539   & 23 46 27.36 & 23 19.46 &   &  &	         & & & \nl
MGC2350+2331   & 23 47 43.12 & 15 19.61 &  &  &	          1.693   &  0.001  &  Q   &  Ly$\alpha$, SiIV, CIV, HeII, \ciii, MgII  \nl
MGC2356+3840   & 23 54 26.73 & 23 33.25 & 18.5  & 0.2 &	  0.2281  &  0.0003 & E &  (HK, G, Mg, Na)  \nl
\hline
\enddata
\small
\tablecomments{
See Table \ref{s1id} for comments and definitions of object types. 
}
\label{s2id}
\end{deluxetable}

\clearpage

\begin{deluxetable}{lccccllll}
\scriptsize
\tablecaption{Sample 3 (MG  50--100 mJy)}
\tablehead{
\colhead{Object}  &  
\colhead{$\alpha$ (B1950)} &    
\colhead{$\delta$ (B1950)} & 
\colhead{$I$} & \colhead{$\sigma_I$} & \colhead{$z$} & \colhead{$\sigma_z$} & 
\multicolumn{2}{l}{\mbox{type \hspace{1cm} detected lines}}}
\startdata
MG0803+3055  & 08 00 24.04 & 31 05 04.57 & 19.6 & 0.2  &  & & & \nl
MG0809+3122  & 08 06 05.02 & 31 31 12.00 & 15.7 & 0.1  &   0.220 &  0.001   & b &   (HK, G, Mg, Na)  \nl
MG0814+2809  & 08 11 55.85 & 28 18 47.00 & 20.5 & 0.2  &   (0.138) & 0.006   & L &  (HK, H$\delta$, Na), OIII, H$\alpha$  \nl
MG0828+2919  & 08 25 05.42 & 29 30 17.01 & 18.5 & 0.1  &   2.322   &  0.005   &  Q   &  OVI, Ly$\alpha$, CIV, MgVII  \nl
MG0854+3009  & 08 51 31.15 & 30 21 24.85 & 21.6 & 0.3 	  &  & &D & \nl
MG0909+2911  & 09 06 16.86 & 29 23 40.33 & 20.2 & 0.2  &  & &D & \nl
MG0920+2755  & 09 17 30.79 & 28 08 38.00 & 23.1 & 1.0 	  &  & &D & \nl
MG0923+3059  & 09 20 07.97 & 31 12 18.00 & 17.2 & 0.1  &   0.6292  &  0.0006  &  Q   &  MgVII, MgII, NeV, HeI, H$\gamma$, H$\beta$  \nl
MG0926+2758  & 09 23 49.16 & 28 11 23.00 &  &  	  &  & & & \nl
MG0932+2837  & 09 29 18.29 & 28 50 47.00 & 17.6 & 0.1   &  0.3033  &  0.0002  & E &  (HK, H$\delta$, G, H$\beta$, Mg, CaFe, Na)  \nl
MG0933+2844  & 09 30 41.39 & 28 58 52.00 & 18.3 & 0.2  &   3.428   &  0.002   &  N  &  Ly$\beta$, Ly$\alpha$, SiIV, CIV  \nl
MG0940+3015  & 09 37 22.49 & 30 28 47.00 & 17.8 & 0.1   &  1.594   &  0.002   &  Q   &  SiIV, CIV, HeII, \ciii, CII, MgII  \nl
MG1013+3042  & 10 10 15.13 & 30 58 25.00 & 18.4 & 0.1 	  &  & & & \nl
MG1019+3037  & 10 16 29.21 & 30 52 45.00 & 20.3 & 0.2  &   1.342   &  0.002   &  Q   &  CIV, HeII, \ciii, MgVII, MgII  \nl
MG1023+2856  & 10 20 34.89 & 29 12 02.00 & 17.3 & 0.1 &   & & & \nl
MG1028+3107  & 10 25 27.83 & 31 22 53.99 & 17.6 & 0.1 &    0.2403  &  0.0005  & E &  (HK, G, Mg, CaFe, Na), MgII, H$\alpha$  \nl
MG1044+2958  & 10 41 19.77 & 30 14 46.00 & 18.0 & 0.1  &   2.981   &  0.001   &  Q   &  OVI, SIV, Ly$\alpha$, SiIV, CIV  \nl
MG1045+3143  & 10 42 36.19 & 31 58 18.00 & 18.5 & 0.2 &    3.230   &  0.005   &  N   &  SIV, OVI, Ly$\alpha$, SiIV, CIV  \nl
MG1106+3000  & 11 03 41.30 & 30 16 58.00 &  &  	 &   & & & \nl
MG1111+2841$^\dagger$  & 11 08 31.45 & 28 58 05.00 &  &    	 &   0.02937   & 0.00003         & E & \nl
MG1112+2844  & 11 10 05.92 & 29 00 03.85 &  &  	 &   & & & \nl
MG1137+2935  & 11 34 43.15 & 29 52 15.00 & 17.7 & 0.2 &    2.644   &  0.001   &  N  &  OVI, Ly$\alpha$, SiIV, CIV, HeII, \ciii  \nl
MG1142+2855  & 11 40 17.63 & 29 11 27.00 &  &  	 &    0.0974  &  0.0002  & E &  (HK, H$\delta$, G, H$\beta$, Mg, CaFe, Na)  \nl
MG1145+2800  & 11 43 11.88 & 28 17 56.00 & 19.7 & 0.1 &   & & & \nl
MG1146+2845  & 11 44 11.73 & 29 01 22.00 &  &  	 &   & & & \nl
MG1202+2756$^\dagger$  & 12 00 00.50 & 28 13 07.00 &  &    	   &  0.672    &          &  Q   & \nl
MG1213+2812  & 12 10 57.91 & 28 28 31.00 &  &  	 &   & & & \nl
MG1215+2750  & 12 13 18.26 & 28 06 16.00 & 16.6 & 0.2 &    0.1034  &  0.0001  & E &  (HK, G, H$\beta$, Mg, CaFe, Na)  \nl
MG1301+2822$^\dagger$  & 12 58 55.76 & 28 37 44.00 &  &    	    &  1.373    &          &  Q   & \nl
MG1310+2925$^\dagger$  & 13 07 43.24 & 29 42 15.00 & 18.0 & 0.2 & 1.21  & & Q  & \nl
MG1312+3113  & 13 10 27.54 & 31 28 53.00 & 16.6 & 0.1 &    1.0533  &  0.0009  &  Q   &  \ciii, MgII, NeV  \nl
MG1334+3043  & 13 32 04.52 & 30 59 32.00 & 15.4 & 0.1 &    1.352   &  0.001   &  N  &  SiIV, CIV, NIII, \ciii, MgII  \nl
MG1340+3009  & 13 38 24.62 & 30 23 43.00 & 17.4 & 0.1 &   & & & \nl
MG1342+2828$^\dagger$  & 13 40 36.25 & 28 43 10.00 &  &    	    &  1.037    &          &  Q   & \nl
MG1346+2900  & 13 44 20.21 & 29 15 40.00 &  &  	 &   & & & \nl
MG1347+2836$^\ddagger$  & 13 45 34.25 & 28 51 25.00 & 13.5 & 0.1 &   & &D & \nl
MG1353+2933  & 13 51 40.75 & 29 47 50.00 & 20.2 & 0.2 &   & &D & \nl
MG1354+3139$^\dagger$  & 13 51 51.20 & 31 53 45.00 &  &     &  1.326   &          &  Q   & \nl
MG1355+3023$^\dagger$  & 13 53 26.22 & 30 38 51.00 &  &  	    &  1.018    &          &  Q   & \nl
MG1356+2918  & 13 54 37.00 & 29 32 55.00 & 18.7 & 0.1    &  3.244   &  0.005   &  N  &  Ly$\alpha$, SiIV, CIV  \nl
MG1400+2918  & 13 57 53.82 & 29 32 57.00 &  &  	  & & & & \nl
MG1406+2930  & 14 03 56.65 & 29 45 58.00 &  &  	  &  & & & \nl
MG1415+2823  & 14 13 23.64 & 28 37 14.00 & 16.6 & 0.1   &   0.2243  &  0.0003  & E &  (HK, G, H$\beta$, Mg, Na)  \nl
MG1437+3119$^\dagger$  & 14 35 31.49 & 31 31 57.00 &  &            &  1.366    &          &  Q   & \nl
MG1438+3001  & 14 35 49.42 & 30 15 03.00 & 16.9 & 0.2  &    0.2316  &  0.0003  & E &  (G, H$\beta$, Mg, CaFe, Na), MgII, NeV  \nl
\hline
\enddata
\small
\tablecomments{
See Table \ref{s1id} for comments and definitions of object types. 
A $\ddagger$ indicates likely contamination of the magnitude by a
foreground star. 
}
\label{s3id}
\end{deluxetable}

\clearpage

\begin{deluxetable}{lccllll}
\scriptsize
\tablecaption{Serendipitous Objects in JVAS (200-250 mJy)}
\tablehead{
  \colhead{Object} &\colhead{$\alpha$ (B1950)} &\colhead{$\delta$ (B1950)}  
     &\colhead{$z$} &\colhead{$\sigma_z$} &\colhead{type} &\colhead{lines}}

\startdata
0707+424 &07 10 44.3 &42 20 55.0 &  1.1645  &  0.0003 &  Q   &  \ciii, MgII  \nl
0718+374 &07 22 01.6 &37 22 28.6 &  1.629   &  0.001  &  Q   &  CIV, \ciii, MgII  \nl
0806+350 &08 09 38.9 &34 55 37.3 &  0.0823  &  0.0002 & L &  (HK, G, H$\beta$, Mg, CaFe, Na), H$\alpha$  \nl
0932+367 &09 35 31.8 &36 33 17.6 &  2.852   &  0.003  &  Q &  Ly$\alpha$, CIV, HeII, \ciii  \nl
1035+430 &10 38 18.2 &42 44 42.8 &  0.3055  &  0.0002 & L &  (HK, H$\delta$, G), OII, H$\beta$, OIII \nl
\hline
\enddata
\small
\tablecomments{ 
See Table \ref{s1id} for comments and definitions of object types. 
}
\label{serendip}
\end{deluxetable}

\clearpage

\def\eph{\hphantom{1.60}}
\begin{deluxetable}{lrrcc}
\footnotesize
\tablewidth{0pt}
\tablecaption{Limits on Flat Cosmological Models}
\tablehead{ 
  \colhead{Model} &$\ln L$ 
&\colhead{max} &\colhead{1-$\sigma$} &\colhead{2-$\sigma$} 
   }
\startdata
\hline
OPT              &$ -34.98 $  &$ 1.24 $  & $ 0.65 < \Omega_0 < \eph $  & $ 0.31 < \Omega_0 < \eph $  \nl 
OPT--S           &$ -35.17 $  &$ 1.59 $  & $ 0.94 < \Omega_0 < \eph $  & $ 0.55 < \Omega_0 < \eph $  \nl 
RAD--A           &$ -11.05 $  &$ 0.86 $  & $ 0.50 < \Omega_0 < 1.47 $  & $ 0.29 < \Omega_0 < \eph $  \nl 
RAD--B           &$ -11.05 $  &$ 1.01 $  & $ 0.60 < \Omega_0 < 1.69 $  & $ 0.36 < \Omega_0 < \eph $  \nl 
RAD--C           &$ -11.05 $  &$ 0.80 $  & $ 0.47 < \Omega_0 < 1.38 $  & $ 0.27 < \Omega_0 < \eph $  \nl 
RAD--A+OPT       &$ -46.55 $  &$ 1.07 $  & $ 0.66 < \Omega_0 < 1.71 $  & $ 0.39 < \Omega_0 < \eph $  \nl 
RAD--B+OPT       &$ -46.32 $  &$ 1.15 $  & $ 0.72 < \Omega_0 < 1.81 $  & $ 0.44 < \Omega_0 < \eph $  \nl 
RAD--C+OPT       &$ -46.66 $  &$ 1.04 $  & $ 0.64 < \Omega_0 < 1.66 $  & $ 0.38 < \Omega_0 < \eph $  \nl 
RAD--A+OPT--S    &$ -47.01 $  &$ 1.28 $  & $ 0.84 < \Omega_0 < 1.95 $  & $ 0.55 < \Omega_0 < \eph $  \nl 
RAD--B+OPT--S    &$ -46.71 $  &$ 1.35 $  & $ 0.89 < \Omega_0 < \eph $  & $ 0.58 < \Omega_0 < \eph $  \nl 
RAD--C+OPT--S    &$ -47.15 $  &$ 1.26 $  & $ 0.82 < \Omega_0 < 1.92 $  & $ 0.54 < \Omega_0 < \eph $  \nl 
RAD--C+OPT--a01  &$ -46.42 $  &$ 0.97 $  & $ 0.59 < \Omega_0 < 1.56 $  & $ 0.35 < \Omega_0 < \eph $  \nl 
RAD--C+OPT--a02  &$ -46.22 $  &$ 0.91 $  & $ 0.56 < \Omega_0 < 1.46 $  & $ 0.33 < \Omega_0 < \eph $  \nl 
RAD--C+OPT--a03  &$ -46.06 $  &$ 0.85 $  & $ 0.52 < \Omega_0 < 1.37 $  & $ 0.30 < \Omega_0 < \eph $  \nl 
RAD--C+OPT--a04  &$ -45.95 $  &$ 0.80 $  & $ 0.49 < \Omega_0 < 1.29 $  & $ 0.29 < \Omega_0 < \eph $  \nl 
RAD--C+OPT--a05  &$ -45.88 $  &$ 0.75 $  & $ 0.46 < \Omega_0 < 1.21 $  & $ 0.27 < \Omega_0 < 1.93 $  \nl 
RAD--C+OPT--a06  &$ -45.87 $  &$ 0.71 $  & $ 0.43 < \Omega_0 < 1.15 $  & $ 0.26 < \Omega_0 < 1.83 $  \nl 
RAD--C+OPT--a07  &$ -45.91 $  &$ 0.68 $  & $ 0.41 < \Omega_0 < 1.09 $  & $ 0.24 < \Omega_0 < 1.74 $  \nl 
RAD--C+OPT--a08  &$ -45.99 $  &$ 0.64 $  & $ 0.39 < \Omega_0 < 1.03 $  & $ 0.23 < \Omega_0 < 1.66 $  \nl 
RAD--C+OPT--a09  &$ -46.15 $  &$ 0.62 $  & $ 0.37 < \Omega_0 < 0.99 $  & $ 0.22 < \Omega_0 < 1.58 $  \nl 
RAD--C+OPT--a10  &$ -46.37 $  &$ 0.59 $  & $ 0.36 < \Omega_0 < 0.95 $  & $ 0.20 < \Omega_0 < 1.52 $  \nl 
RAD--C+OPT--f01  &$ -48.78 $  &$ 0.49 $  & $ 0.29 < \Omega_0 < 0.79 $  & $ 0.18 < \Omega_0 < 1.28 $  \nl 
RAD--C+OPT--f02  &$ -46.85 $  &$ 0.55 $  & $ 0.34 < \Omega_0 < 0.89 $  & $ 0.19 < \Omega_0 < 1.44 $  \nl 
RAD--C+OPT--f03  &$ -46.18 $  &$ 0.62 $  & $ 0.38 < \Omega_0 < 1.00 $  & $ 0.22 < \Omega_0 < 1.60 $  \nl 
RAD--C+OPT--f04  &$ -45.95 $  &$ 0.68 $  & $ 0.41 < \Omega_0 < 1.10 $  & $ 0.25 < \Omega_0 < 1.76 $  \nl 
RAD--C+OPT--f05  &$ -45.91 $  &$ 0.74 $  & $ 0.45 < \Omega_0 < 1.19 $  & $ 0.27 < \Omega_0 < 1.91 $  \nl 
RAD--C+OPT--f06  &$ -45.98 $  &$ 0.81 $  & $ 0.49 < \Omega_0 < 1.29 $  & $ 0.29 < \Omega_0 < \eph $  \nl 
RAD--C+OPT--f07  &$ -46.11 $  &$ 0.86 $  & $ 0.53 < \Omega_0 < 1.39 $  & $ 0.31 < \Omega_0 < \eph $  \nl 
RAD--C+OPT--f08  &$ -46.28 $  &$ 0.92 $  & $ 0.56 < \Omega_0 < 1.48 $  & $ 0.33 < \Omega_0 < \eph $  \nl 
RAD--C+OPT--f09  &$ -46.46 $  &$ 0.98 $  & $ 0.60 < \Omega_0 < 1.57 $  & $ 0.36 < \Omega_0 < \eph $  \nl 
\hline
Type Ia Supernova &           &$ 0.94 $  & $ 0.66 < \Omega_0 < 1.28 $  & $ 0.51 < \Omega_0 < \eph $  \nl
\hline
\enddata
\tablecomments{An empty entry means that the statistical limit was not reached at the edge of the range 
 $0 < \Omega_0 < 2$ and $-1 < \lambda_0 < 1$.  ``RAD-A (B,C)" designates the radio data with completeness
 model A (B, C), ``OPT'' and ``OPT-S'' designate the optical data either with or without spiral galaxy lenses,
 $-axx$ indicates a mean magnitude change of $\dm = xx$, and $-fxx$ indicates an optical completeness of
 $f=xx$.  The Perlmutter et al. (1997) results for Type Ia supernovae
 are also shown, with a blank entry indicating that the limits was not given. }
\label{models1}
\end{deluxetable}

\clearpage

\begin{deluxetable}{lrrcc}
\footnotesize
\tablewidth{0pt}
\tablecaption{Limits on $\lambda_0=0$ Cosmological Models}
\tablehead{ 
  \colhead{Model} &$\ln L$ &\colhead{max} &\colhead{1-$\sigma$} &\colhead{90\% confid}d
   }
\startdata
\hline
OPT              &$ -35.00 $  &$ 1.38 $  & $ 0.30 < \Omega_0 < \eph $  & $ \eph < \Omega_0 < \eph $  \nl 
OPT--S           &$ -39.58 $  &$ 1.95 $  & $ \eph < \Omega_0 < \eph $  & $ \eph < \Omega_0 < \eph $  \nl 
RAD--A           &$ -11.06 $  &$ 1.12 $  & $ 0.32 < \Omega_0 < \eph $  & $ \eph < \Omega_0 < \eph $  \nl 
RAD--B           &$ -10.97 $  &$ 1.45 $  & $ 0.49 < \Omega_0 < \eph $  & $ 0.13 < \Omega_0 < \eph $  \nl 
RAD--C           &$ -11.12 $  &$ 1.09 $  & $ 0.24 < \Omega_0 < \eph $  & $ \eph < \Omega_0 < \eph $  \nl 
RAD--A+OPT       &$ -46.46 $  &$ 1.44 $  & $ 0.55 < \Omega_0 < \eph $  & $ 0.19 < \Omega_0 < \eph $  \nl 
RAD--B+OPT       &$ -46.21 $  &$ 1.53 $  & $ 0.65 < \Omega_0 < \eph $  & $ 0.27 < \Omega_0 < \eph $  \nl 
RAD--C+OPT       &$ -46.59 $  &$ 1.42 $  & $ 0.51 < \Omega_0 < \eph $  & $ 0.16 < \Omega_0 < \eph $  \nl 
RAD--A+OPT--S    &$ -46.82 $  &$ 1.77 $  & $ 0.88 < \Omega_0 < \eph $  & $ 0.46 < \Omega_0 < \eph $  \nl 
RAD--B+OPT--S    &$ -46.50 $  &$ 1.92 $  & $ 0.97 < \Omega_0 < \eph $  & $ 0.54 < \Omega_0 < \eph $  \nl 
RAD--C+OPT--S    &$ -46.97 $  &$ 1.74 $  & $ 0.84 < \Omega_0 < \eph $  & $ 0.43 < \Omega_0 < \eph $  \nl 
RAD--C+OPT--a01  &$ -46.38 $  &$ 1.14 $  & $ 0.43 < \Omega_0 < \eph $  & $ 0.11 < \Omega_0 < \eph $  \nl 
RAD--C+OPT--a02  &$ -46.21 $  &$ 1.11 $  & $ 0.36 < \Omega_0 < \eph $  & $ 0.06 < \Omega_0 < \eph $  \nl 
RAD--C+OPT--a03  &$ -46.11 $  &$ 1.07 $  & $ 0.29 < \Omega_0 < \eph $  & $ 0.01 < \Omega_0 < \eph $  \nl 
RAD--C+OPT--a04  &$ -46.02 $  &$ 0.80 $  & $ 0.22 < \Omega_0 < 1.86 $  & $ \eph < \Omega_0 < \eph $  \nl 
RAD--C+OPT--a05  &$ -45.99 $  &$ 0.77 $  & $ 0.17 < \Omega_0 < 1.71 $  & $ \eph < \Omega_0 < \eph $  \nl 
RAD--C+OPT--a06  &$ -46.03 $  &$ 0.72 $  & $ 0.12 < \Omega_0 < 1.59 $  & $ \eph < \Omega_0 < \eph $  \nl 
RAD--C+OPT--a07  &$ -46.10 $  &$ 0.59 $  & $ 0.08 < \Omega_0 < 1.47 $  & $ \eph < \Omega_0 < \eph $  \nl 
RAD--C+OPT--a08  &$ -46.22 $  &$ 0.45 $  & $ 0.04 < \Omega_0 < 1.34 $  & $ \eph < \Omega_0 < \eph $  \nl 
RAD--C+OPT--a09  &$ -46.38 $  &$ 0.43 $  & $ 0.01 < \Omega_0 < 1.23 $  & $ \eph < \Omega_0 < \eph $  \nl 
RAD--C+OPT--a10  &$ -46.63 $  &$ 0.42 $  & $ \eph < \Omega_0 < 1.16 $  & $ \eph < \Omega_0 < 1.90 $  \nl 
RAD--C+OPT--f01  &$ -49.15 $  &$ 0.21 $  & $ \eph < \Omega_0 < 0.82 $  & $ \eph < \Omega_0 < 1.45 $  \nl 
RAD--C+OPT--f02  &$ -47.15 $  &$ 0.39 $  & $ \eph < \Omega_0 < 1.04 $  & $ \eph < \Omega_0 < 1.76 $  \nl 
RAD--C+OPT--f03  &$ -46.41 $  &$ 0.43 $  & $ 0.01 < \Omega_0 < 1.25 $  & $ \eph < \Omega_0 < \eph $  \nl 
RAD--C+OPT--f04  &$ -46.13 $  &$ 0.60 $  & $ 0.08 < \Omega_0 < 1.49 $  & $ \eph < \Omega_0 < \eph $  \nl 
RAD--C+OPT--f05  &$ -46.03 $  &$ 0.77 $  & $ 0.15 < \Omega_0 < 1.68 $  & $ \eph < \Omega_0 < \eph $  \nl 
RAD--C+OPT--f06  &$ -46.05 $  &$ 0.81 $  & $ 0.22 < \Omega_0 < 1.88 $  & $ \eph < \Omega_0 < \eph $  \nl 
RAD--C+OPT--f07  &$ -46.15 $  &$ 1.08 $  & $ 0.31 < \Omega_0 < \eph $  & $ 0.02 < \Omega_0 < \eph $  \nl 
RAD--C+OPT--f08  &$ -46.27 $  &$ 1.11 $  & $ 0.37 < \Omega_0 < \eph $  & $ 0.07 < \Omega_0 < \eph $  \nl 
RAD--C+OPT--f09  &$ -46.42 $  &$ 1.14 $  & $ 0.44 < \Omega_0 < \eph $  & $ 0.12 < \Omega_0 < \eph $  \nl 
\hline
Type Ia Supernova &           &$ 0.88 $  & $ 0.28 < \Omega_0 < 1.57 $  & $ \eph < \Omega_0 < \eph $  \nl
\hline
\enddata
\tablecomments{See Table \ref{models1} for comments and model definitions.}
\label{models2}
\end{deluxetable}

\end{document}